\documentclass[11pt]{article}
\usepackage{epsfig}
\usepackage[usenames]{color}
\usepackage{graphicx}
\usepackage{amsmath}
\def\la{\langle}\def\ra{\rangle}
\def\be{\begin{eqnarray}}\def\ee{\end{eqnarray}}
\def\lsim{\mathrel{\rlap{\lower3pt\hbox{\hskip1pt$\sim$}}
     \raise1pt\hbox{$<$}}} %less than or approx. symbol
\def\gsim{\mathrel{\rlap{\lower3pt\hbox{\hskip1pt$\sim$}}
     \raise1pt\hbox{$>$}}} %greater than or approx. symbol
\def\le{ \begin{array}{ll}}\def\re{\end{array}}

\def\lear{ \left( \begin{array}{cc}}\def\rear{\end{array} \right)}

\def\le{ \left( \begin{array}{cc}}\def\re{\end{array} \right)}

\def\bi{\bibitem}

\def\eft-hls{{\it EFT}$_{\rm bsHLS}$}
\def\skyrmion-hls{{\it Skyrmion}$_{\rm sHLS}$}
\def\del{\partial}

\renewcommand{\thefootnote}{\fnsymbol{footnote}}
\begin{document}

\centerline{\Large \bf The ``Folk Theorem" on Effective Field Theory:}

\centerline{\Large \bf How Does It Fare in Nuclear Physics? }
\vskip 0.5cm
\begin{center}

Mannque Rho$^a$\footnote{\sf e-mail: mannque.rho@ipht.fr}

\vskip 0.3cm

{\em $^a$Institut de Physique Th\'eorique, CEA Saclay, 91191 Gif-sur-Yvette c\'edex, France }

\vskip 0.3cm
{\ (\today)}
\end{center}

\centerline{ \bf ABSTRACT}
\vskip 0.2cm
This is a brief history of what I consider as very important, some of which  truly seminal, contributions made by young Korean nuclear theorists, mostly graduate students working on PhD thesis in 1990s and  early 2000s,  to nuclear effective field theory, nowadays heralded as the first-principle approach to nuclear physics.  The theoretical framework employed is an effective field theory anchored on  a {\it single} scale-invariant hidden local symmetric Lagrangian constructed in the spirit  of Weinberg's ``Folk Theorem" on effective field theory.  The problems addressed  are the high-precision calculations on the thermal $np$ capture, the solar $pp$ fusion process, the solar  $hep$ process -- John Bahcall's challenge to nuclear theorists -- and  the quenching of $g_A$ in giant Gamow-Teller resonances and the whopping enhancement of first-forbidden beta transitions relevant in astrophysical processes.  Extending adventurously the strategy to a wild uncharted domain in which a systematic implementation of the ``theorem" is far from obvious, the {\it same} effective Lagrangian is applied to the structure of compact stars. A surprising, unexpected, result on the properties of massive stars, totally different from what has been obtained up to day in the literature,  is predicted, such as the precocious onset of conformal sound velocity together with a hint for the possible emergence in dense matter of  hidden symmetries such as scale symmetry and hidden local symmetry.

%\tableofcontents
\noindent
\setcounter{footnote}{0}
\renewcommand{\thefootnote}{\arabic{footnote}}
\vskip 1.0cm
\section{Introduction}
With supersymmetry still invisible at the  super-accelerator energies at CERN, and consequently string theory, the would-be ``Theory of Everything" (TOE), in -- perhaps temporary -- doldrums, it has become a vogue to resort to effective field theories to make further progress, ranging  from particle physics at the smallest scale all the way  to cosmology dealing with large scale universe. To some extent this may be more of a desperation, given no other alternatives available, but the currently favored view is to think that independently of whether there is ultimately a TOE,  going the route of effective field theories would be an extremely fruitful way to unravel Nature.  This attitude applies even to the case where a potential ``fundamental theory" exists, such as for instance QED for condensed matter physics and QCD for nuclear physics. Although both are ``effective theories" from the point of view of TOE, they can be considered a complete theory within the domain relevant to these subsystems.

In this article I address the second matter in connection with nuclear physics.

There is little doubt that QCD is the correct fundamental theory for strong interactions.  With nuclear interactions involving strong interactions,  QCD therefore must be able to explain {\it everything} that's going on in nuclear physics. Just to cite a few examples, QCD should explain how the first  excited $0^+$ state at 6.06 MeV of $^{16}$O  on top of the spherical ground state is deformed, the structure of the rotational band of the famous Hoyle state in $^{12}$C, the long life-time of $^{14}$C responsible for the carbon-14 dating etc.  These apparently complex and intriguing phenomena have been fairly sell explained in various models developed in the field since a long time in nuclear physics circle without any recourse to the QCD degrees of freedom,  quarks and gluons. Attempts are being made to go from QCD proper to nuclear processes using lattice techniques, and there have been some progresses~\cite{savage}, in particular, in calculating  baryon-baryon potentials, essential for many-body systems,  electroweak response functions for light nuclei and so on. So far whatever has been feasible for reliable calculations shows that QCD \`a la lattice does give post-dictions satisfactorily and indicate that ultimately such first-principle calculations will achieve to explain many, if not all, of nuclear processes that have been calculated with some accuracy, including perhaps what's mentioned above, in the standard nuclear physics approach (SNPA for short). The only thing is that it will take a very very long time.

Now what about the highly compressed hadronic matter considered to be present inside massive compact stars?

Due to the famous sign problem, it is not at all clear when and how this problem will be solved by lattice techniques. Although at low density below nuclear matter, various approximate methods could be employed, the high density involved in the stars cannot be accessed at all. There are no other known techniques in QCD that can unravel what happens in the $\sim 2$-solar mass stars recently determined~\cite{massive-stars}. At superhigh density, way above the pertinent density,  perturbative QCD is perhaps applicable, but this perturbative regime is most likely irrelevant to the problem. So it looks that even if there is a well-defined ``theory of everything" within the nuclear domain, it is not directly relevant.

How does one go about talking about the deformed $0^+$ state in $^{16}$O, the Hoye state etc. on the one hand and on the other hand, unravel what happens in dense compact stars and possibly gravity waves coming from merging neutron stars?

This is where effective field theories come in. I would like to recount what effective field theory is all about and how it figures in nuclear physics and what seminal contributions several Korean nuclear theorists have made to the early development in the field.

Let me start with what is called ``Folk Theorem" on effective quantum field theory.
\section{The Folk Theorem}

What is quantum field theory and how does effective field theory (EFT for short) work? This question was raised some time ago and has since then been answered by Weinberg in terms of a  ``theorem"~\cite{weinberg-festschrift,folk-theorem}. It is classified as a ``Folk Theorem," not a rigorous physical or mathematical  theorem, because as explained by Weinberg, it is not a bona-fide theorem in the sense it can be given a rigorous proof approved by axiomatic field theorists. This is the reason why I put the quotation mark.  Nonetheless nowadays it figures legitimately in field theory textbooks~\cite{weinberg-book2}. This note is an offer of support to that theorem from nuclear physics. It is in my mind the most concrete and perhaps the most successful case of how it works.

The ``Folk Theorem" (FT for short) states: ``If one writes down the most general possible Lagrangian, including all terms consistent with assumed symmetry principles, and then calculates matrix elements with this Lagrangian to any given order of  perturbation theory, the result will simply be the most general possible S-matrix consistent with perturbative unitarity,  analyticity, cluster decomposition, and the assumed symmetry  properties." Though it does not enjoy a rigorous proof,  one cannot see how it can go wrong

The point is that this theorem should apply to {\it all} interactions in Nature even if there existed  a ``Theory of Everything."  The most prominent cases are seen in condensed matter physics. In strongly correlated condense matter systems,  a large variety of effective field theories  are uncovering  a wealth of fascinating phenomena, exposing emergent symmetries with no visible connections to  QED, the ``fundamental theory" for condensed matter systems.  In particle physics, the Standard Model is a preeminent case of effective field theory in search of more fundamental theory or  ``ultimate theory," which can be solved to a great accuracy to confront Nature~\cite{eft-particlephysics}. Unlike in QED, however, QCD, the theory for strong interactions, cannot yet be solved in the nonperturbative regime and hence cannot have a direct access to nuclear phenomena. Thus effective field theory is  the only  approach available in nuclear physics. And it has proven to be highly successful.

There are two ways the FT could play a crucial role in nuclear physics. One way, Class A,  is that it can be exploited for high precision calculations for the nuclear processes that are accurately known empirically but cannot be calculated reliably by available methods. Here effective field theories representing QCD as faithfully as feasible in the sense of the FT could give highly precise results. This will be a  post-diction, but it can also lead to spin-off predictions related to them.  The other, Class B,  is that it can be exploited for making genuine predictions of what could happen in the domain that QCD proper cannot access within a foreseeable  future. For the class A, I will describe certain response functions in light nuclear systems and for  the class B, what takes place in highly dense baryonic matter as what's expected to be found in compact stars for which the lattice QCD  is currently powerless.

In this paper, it will be argued that this success in nuclear physics in the class A could be taken as  a  ``folk proof" of the FT. In my view, this offers the most solid case of how the theorem works in strong interaction physics. In fact the initial development dates way back to 1960's with Skyrme's pioneering work on unified theory of mesons and baryons~\cite{Skyrme-Model} and his energy density functional approach (known in the literature Skyrme potential)  in nuclear physics~\cite{skyrme-potential}.\footnote{The two ideas brought out by Skyrme turn out to be connected in an intricate and involved way. This matter is however out of the scope of this note. See \cite{MaRho} for discussions on this matter.}  It  has then evolved with the recognition of chiral symmetry in nuclear dynamics via current algebras in 1970's before the advent of QCD and then chiral perturbation theory in 1990's for nuclear forces and nuclear response functions.

Here  Korean nuclear theorists made seminal contributions. It is one of the objectives of this paper to bring this to the attention of the Korean physics community.

This series of developments in nuclear physics is aptly captured by Weinberg's words~\cite{folk-theorem}: ``The use of effective quantum field theories has been extended more recently to nuclear physics, where although nucleons are not soft they never get far from their mass shell, and for that reason can be also treated by similar methods as the soft pions. Nuclear physicists have adopted this point of view, and I gather that they are happy about using this new language because it allows one to show in a fairly convincing way that what they've been doing all along  ....  is the correct first step in a consistent approximation scheme." Let me call this observation ``Corollary" to the FT.

I first illustrate, with a few selected cases,  how the ``proof" works out in nuclear processes.  In the coming years,  the folk theorem will be more quantitatively verified in nuclear physics in the vicinity of  normal nuclear matter density $n_0$ or slightly above as higher-order corrections, in principle systematically doable,  are calculated with more powerful numerical techniques within the well-defined effective field theory scheme and more and higher-accuracy experimental data become available. Given an EFT framework accurately formulated along the FT, what is needed is then high-powered numerical techniques to solve the {\it given} well-defined many-body problems.  Such techniques have been actively developed -- and are being developed further -- outside of the context of the EFT formalism, and the recent development renders such techniques implementable to the EFT strategy.  Some notable examples, just to name a few, are the {\it ab initio} methods with or without core~\cite{ab initio},  Green's function methods~\cite{greens function}, quantum Monte Carlo~\cite{QMC}, renormalization-group strategy~\cite{RG} etc.

The next objective is to go beyond what's established by the FT under normal conditions to the domain of nuclear physics for which very little is established, both theoretically and experimentally.  One particularly challenging -- and very poorly understood -- domain is the high density regime that is considered to be relevant for massive compact stars.  QCD, with the lattice technique hampered by the sign problem, cannot access such conditions. Therefore effective field theory is the only tool that offers the possibility.  Given the paucity of experimental information and trustful model-independent tools currently available, this endeavor is unquestionably exploratory. It involves assumptions and simplifications, with full of holes in the reasoning far fetched even from the non-axiomatic FT. What is obtained is admittedly very tentative. However the results are highly promising and represent yet another potential achievement from Korean nuclear theorists.

\section{How Nuclear EFT Fares: Precision Calculation}
Let me start with what I would consider as a ``proof" -- and a very compelling one -- of the FT in nuclear physics.

The current breakthrough in EFT in nuclear physics initially germinated from Weinberg's 1979 article on phenomenological Lagrangians for pion dynamics~\cite{weinberg-festschrift} and got materialized in 1990's in the formulation  from effective chiral Lagrangians of nuclear forces~\cite{weinberg-NEFT,vankolck} and of electroweak responses~\cite{mr91}.  A remarkable progress has been made since then on formulating systematic approaches to wide-ranging nuclear dynamics, ranging from nuclear forces to nuclear structure and confronting nature. There are numerous reviews on the current status of nuclear EFT, among which are some recent ones on nuclear forces~\cite{meissner,machleidt-eft} and on nuclear weak response functions~\cite{riska-schiavilla,epelbaum-axial}. The EFT approach as it stands now is established as the standard  tool to probe strongly-interacting highly correlated matter under normal conditions, say, up to $\sim n_0$ or perhaps slightly above. It is also most likely the {\it only} tool to go beyond the normal towards the extreme conditions met in compact stars and merging of neutron stars giving rise to gravity waves.

To see how this development constitutes a proof of the the FT in nuclear physics, we go back to the pre-QCD period. For this, as stated,  we focus on nuclear response functions to external fields instead of on nuclear forces. This is because while the nuclear currents and spectra are intimately connected to each other in EFT, the nuclear response functions can be more accurately calculated and compared with experiments than the nuclear spectra. It renders precision calculations feasible and meaningful.

That chiral symmetry could play an important role in nuclear physics was recognized already in early 1970's. It was implicit in considerations of pion-mediated nuclear potentials,  but the first realization of its potential impact in nuclear physics was in nuclear response functions to the electroweak external potential.  There was no effective field theory then but just phenomenological Lagrangians built with identified hadronic degrees of freedom considered to be relevant to the kinematic regime concerned. Both the nuclear potentials and response functions were then calculated only -- and by necessity  -- at the tree order. It was in calculating the EW response functions that the soft-pion theorems, applicable at the tree order with the Lagrangian, were applied to pionic exchange currents. Given that the pion-exchange represented dominant contribution, with heavier meson degrees of freedom suppressed by their heavy mass, the soft-pion contribution with constraints from the current algebras were calculated reliably parameter-free.  This represented the first indication that chiral symmetry could figure significantly in nuclear processes~\cite{chemtob-rho}. The further development in this line of reasoning led to the prediction that the  $M1$ matrix element and the weak axial-charge matrix element should receive an important contribution from the soft-pion theorems. It was recognized on the ground of the soft-pion theorems that the axial-charge transitions can have an extremely clean meson-exchange effects~\cite{KDR}.  Indeed the nuclear EFT so formulated led to high precision calculations of electroweak processes in light nuclei, a brief summary of which is given in \cite{precision-eft}.

It is here Korean nuclear theorists first entered. To recount this development, let us consider two processes.

One is on the thermal $np$ capture
\be
n+p\to d+\gamma\label{np}
\ee
and the other is the solar fusion processes
\be
p+p\to d+ e^+ +\nu_e.\label{pp}
\ee

Historically, the process (\ref{np}) was first explained quantitatively two decades ago by Riska and Brown~\cite{riska-brown} who showed that the $\sim 10\%$
discrepancy present at that time between the experimental cross section and the theoretical impulse approximation prediction is eliminated by exchange currents. Riska and Brown computed, using realistic hard-core wavefunctions,  the electroweak exchange current suggested by Chemtob and Rho~\cite{chemtob-rho} coming from current-algebra low-energy theorems. Since the current algebra term is the leading-order contribution in chiral perturbation theory\footnote{We denote by S$\chi$EFT the chiral perturbation theory that involves nucleons and pions only, to be distinguished from chiral effective field theory that includes the light vector mesons, $\rho$ and $\omega$, and also a dilaton scalar, which will be denoted $\sigma$.},  what Riska and Brown calculated is incorporated in the chiral expansion as the leading order term. The first calculation in standard chiral effective field theory (S$\chi$EFT) in the next-to-next-to-leading order (N$^2$LO) in the chiral counting was made by Tae-Sun Park, a graduate student at Seoul National University, giving the result~\cite{np-tsp}
\be
\sigma_{th}=(334 \pm 2) {\rm mb}\label{np-th}
\ee
to be compared with the experiment
\be
\sigma_{exp}=(334.2 \pm 0.5) {\rm mb}.
\ee
The error bar in the theory (\ref{np-th}) represents uncertainty in dealing with the cutoff (short-distance regularization in coordinate space). This is shown, just to highlight the point, in Fig.~\ref{np} (left panel). With the modern refinement in more than two decades after the first S$\chi$EFT calculation, it is very possible that one can do a lot better, eliminating the little holes there might haver been, than what's done in \cite{np-tsp} and reducing even further the theoretical uncertainty.

 \begin{figure}[h]
\begin{center}
\includegraphics[width=6cm]{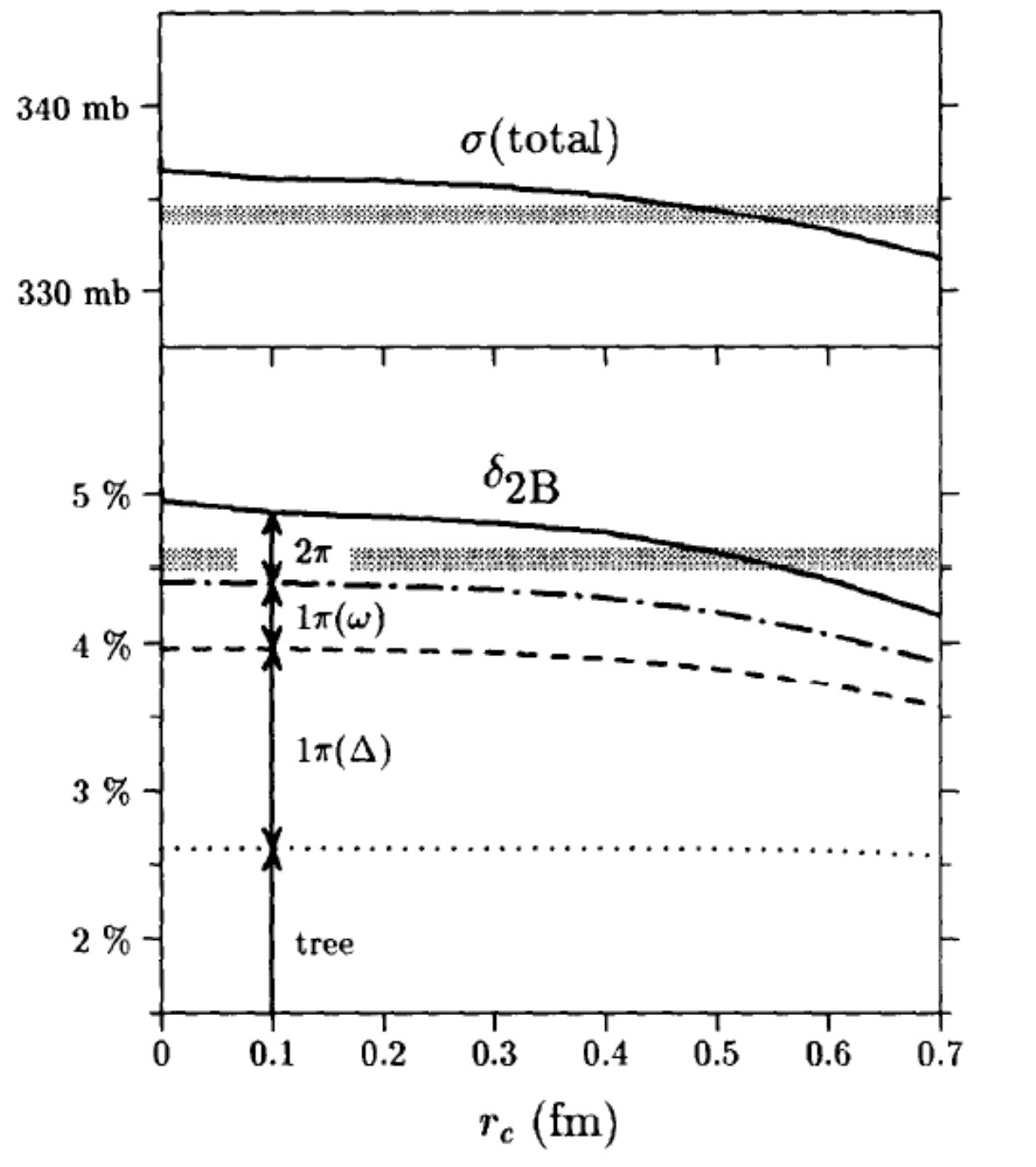}\includegraphics[width=8cm]{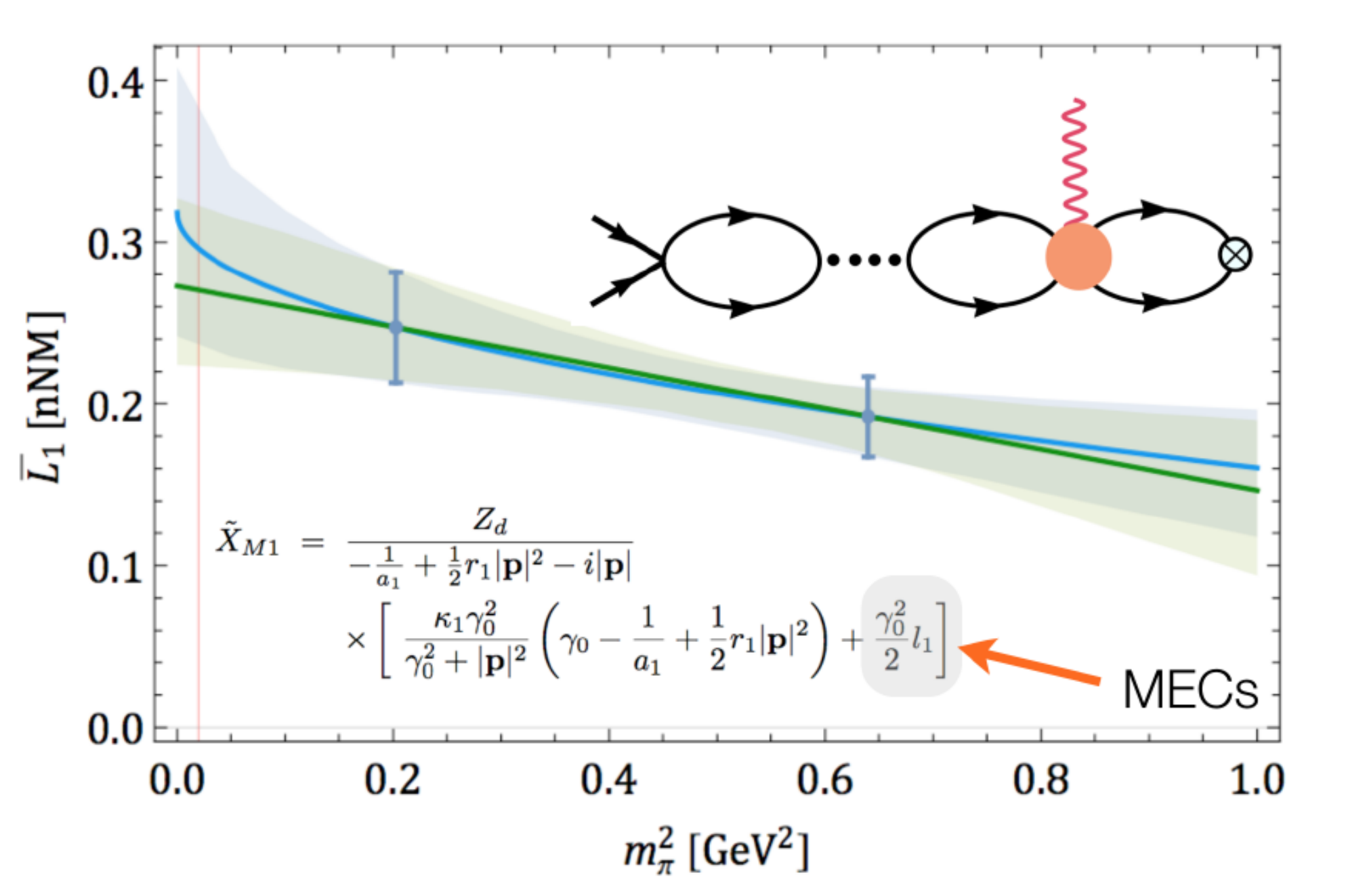}
%\centerline{\epsfig{file=cc.pdf, width=9cm}}
\end{center}
\caption{$n+p\to d+\gamma$. Left panel: S$\chi$EFT calculation for $\sigma$~\cite{np-tsp} vs. $r_c$. Here $\Delta$ and $\omega$ are invoked for the resonance saturation of the constants.
Right panel: Lattice calculation for 2-body correction vs $m_\pi^2$~\cite{np-lattice}. Here the MEC given by  $l_1$, calculated on lattice, contains $m_\pi^2$.}\label{np}
\end{figure}

In the meantime, lattice QCD techniques have advanced well enough to calculate certain light-nuclear processes~\cite{savage}. A recent lattice QCD calculation of this particular process ``measured" with heavy pion mass extrapolated to the physical pion mass using pionless effective field theory which  brings in uncertainty\footnote{Possibly justified for this process, but not for axial charge transitions discussed below.} gave~\cite{np-lattice}
\be
\sigma_{\rm latticeQCD}=334.9.\pm 5.3\label{np-lattice}
\ee
The main uncertainty coming from the pion mass extrapolation is given in Fig.~\ref{np} (right panel). Although apparently unrelated, the errors involved are comparable between the two results. I take this as one of the most beautiful cases that confirm nicely the FT in action in nuclear physics.

Another case that belongs to the class A listed above to which Tae-Sun Park made the {\it pivotal} contribution is the solar $pp$  process (\ref{pp}) and especially the $hep$ process
\be
^{3}{\rm He} +p \to ^{4}{\rm He} + e^+ +\nu_e.\label{hep}
\ee
The processes (\ref{pp}) and (\ref{hep}) are, respectively, the lowest and the highest in the solar neutrino spectra. I will discuss both because the former is very well controlled by S$\chi$EFT, but the latter is highly uncertain, involving higher-order corrections. The problem here is that the two processes involve at the leading order the Gamow-Teller operator. While it contributes rather straightforwardly in the former process,  the latter involves delicate cancellations and hence requires fine-tuned higher-order corrections.

The process (\ref{pp}) was first computed in S$\chi$EFT by Tae-Sun Park in 1998~\cite{pkmr-pp}. Dominated by the Gamow-Teller operator, soft-pions do not contribute, and hence for precision calculation one has to go to higher orders than in the case of the $np$ capture. The exchange current corrections are much smaller than in the case of the $np$ capture. Going to N$^3$LO the result obtained was
\be
S_{pp}(0)=4.05\times 10^{-25} (1 \pm 0.01)\ {\rm MeV\ b}\label{pp1}
\ee
where
\be
S_{pp} (E)=\sigma(E)E e^{2\pi\eta}
\ee
with $\sigma(E)$ the total cross section for the $pp$ fusion and $\eta=m_p\alpha/2p$.  The result (\ref{pp1}) was further improved  to~\cite{tspetal}
\be
S_{pp}(0)=3.91\times 10^{-25} (1\pm 0.008)\ {\rm MeV\ b}.
\ee
The updated accepted value is~\cite{adelbergeretal}
\be
S_{pp}(0)=4.01\times 10^{-25} (1\pm 0.009) {\rm MeV\ b}.\label{adelberger}
\ee
A lattice QCD calculation using a similar pion-mass extrapolation procedure as in the $np$ capture case~\cite{np-lattice} verifies this value, again providing a proof of the FT.
There are other cases that offer equally strong support to how the FT works in nuclear dynamics, e.g.,  $\nu d$ reactions $\nu_e + d\to e^- + p+p$, $\nu + d\to \nu + n+p$, triton beta decay, polarized $np$ capture $\vec{n}+\vec{p}\to d +\gamma$ etc. with tiny theoretical error bars, all more or less postdictions, though highly precise.

When the transition is of Gamow-Teller type, unless there is an accidental cancelation, the single-particle operator dominates. With soft-pion contributions absent, exchange-current corrections are largely suppressed. Thus given accurate wave functions, with the SNPA exchange currents with no systematic (chiral) power counting, as was done in \cite{chemtob-rho}, it is not surprising that the SNPA can also capture rather closely the flux factor (\ref{adelberger}). In fact this is what was found in \cite{schiavilla}. This confirms that the SNPA was going in the right direction, providing a proof to the corollary of the FT. What matters in this case is essentially the nuclear potential -- giving accurate wave functions -- and the sophisticated potential fit to data is as good as, or even better, than what one can get at high $n$ -- in N$^n$LO S$\chi$EFT -- calculations.

Now let me turn to a case where the theoretical situation is completely barren, the $hep$ process (\ref{hep}). This process is the source of the $pp$ chain's most energetic neutrino with an endpoint energy of 18.8 MeV. It has not yet been seen experimentally, with both the Super Kamiokande and SNO collaborations setting only limits on the $hep$ neutrino flux.  It is a formidable task to calculate this process. First, the leading one-body Gamow-Teller operator cannot connect the main (s-wave) component of the initial and final wave functions, second the meson-exchange currents, even with the soft-pion terms suppressed, get enhanced by the p-wave component of the wave functions, and most importantly the single-body and two-body current matrix elements tend to largely cancel. This requires that higher chiral order terms, both in the potential and currents,  be calculated with great accuracy.

Historically from 1952 to 1992, the $hep$ flux $S_{hep}(0)$ ranged in unit of $10^{-20}$ keV barn from 630 to 1.3~\cite{bahcall-krastev}. Given this ``wilderness" in nuclear theory, John Bahcall made a challenge to nuclear theorists in 2000~\cite{bahcall}: ``I do not see anyway at present to determine from experiment or first principles theoretical calculations a relevant, robust upper limit to the $hep$ production cross section (and therefore the $hep$ solar neutrino flux). .. The most important unsolved problem in theoretical nuclear physics related to solar neutrinos is the {\it range} of values allowed by fundamental physics for the $hep$ process cross section."

Bahcall's challenge was first met in 2000 ~\cite{bahcall-challenge}
and then with further precision in 2001~\cite{tsp-hep} by Tae-Sun Park with his collaborators. The result obtained to N$^3$LO in the chiral order is
\be
S_{hep} (0)=(8.6\pm 1.3)\times 10^{-20} {\rm keV\ b}.
\ee
Given the extremely delicate cancelations in the current operators and high sensitivity to the wave functions involved, the theoretical precision is far from what's obtained for  the $pp$ fusion. Nonetheless it is a stunning calculation, not just giving the {\it range} as challenged by Bahcall, but narrowing the order-of-magnitude uncertainty down to a tens of \%, a  prediction with no free parameters. It represented a giant step forward in nuclear theory.  It will of course ultimately be up to Nature to give a verdict on this prediction. A search is going on in the SNP collaboration.

An important point to make here with respect to the FT -- and more appropriately the corollary to it -- is that the standard nuclear physics approach (SNPA) with exchange currents incorporated along the line of \cite{chemtob-rho} with certain constraints consistent with Ward identities~\cite{snpa-hep} gives a result not so numerically different from the  N$^3$LO S$\chi$EFT,
\be
S^{\rm SNPA}_{hep} (0)=(10.1\pm 0.6)\times 10^{-20}\ {\rm keV\ b}.
\ee

The upshot of these results means two things. First, the SNPA -- what nuclear theorists have been doing since very long time in nuclear physics -- was effectively doing the ``right" first step along the line of the FT in QCD. Second, what makes S$\chi$EFT superior in rigor to  SNPA is that corrections to what is calculated can be, in principle, precisely defined, so one can make systematic improvement,  whereas in SNPA, they cannot even be defined other than at the level of soft-pions as in \cite{chemtob-rho}. Therefore it is in practice unknown whether  one can trust what is obtained in SNPAs, however sophisticated it may be -- and even if it may agree with available experiments. This is where the power of effective field theory comes in.

I should mention here that Tae-Sun Park's calculation~\cite{precision-eft} exploits ``accurate" wave functions obtained with what is called ``the most sophisticated or realistic potential" $V$ and the currents $J$ derived to high an $n$ as possible in the N$^n$LO chiral expansion.  (In \cite{precision-eft}, $n_J\sim 3$). Given that $V$ is in principle ``exactly" fit to experimental data, one may consider it as $n=n_V \sim \infty$ in the chiral expansion. This procedure was referred to, provocatively, as ``more effective effective field theory (MEEFT)" or, less provocatively, as EFT$^\ast$ because it exploits the high precision of the potential and the high chiral order of the current  but with $n_J \neq  n_V$. This is a hybrid approach to differentiate it from the ``first-principle" EFT that insists on  $n_V=n_J$. There has been criticism on the ``MEEFT" on the ground that there is a mismatch in the chiral counting between the potential and the current. This criticism is justified since the mismatch can generate ambiguities in implementing the FT to the effective field theory. For instance it could lead to what is known as ``off-shell" ambiguities. This ``in-principle objection" is, however,  moot and insignificant in practice -- except for the ``right-lunatic fringe" bent on rigorous EFT.  In the actual implementation of the FT to ``{\it ab intio} first-principle calculations" in nuclei of mass number greater than 2 or 3,  one necessarily breaks the $n_V=n_J$ constraint. Examples described below illustrate this point. Even for few-body nuclei, there is no known case where the hybridization suffers from the mismatch. One can think of this hybridization as something akin to the scheme dependence in high-order perturbation calculations in QCD  or other gauge theories.

\section{FT in Baryonic Matter}
The next question I raise is: What is EFT in the spirit of the FT when one addresses heavy nuclei, nuclear matter and dense matter? For the systems discussed above, lattice QCD is seen -- and will continue -- to confirm  the validity of the FT. For systems with high density, however, this is no longer feasible. To go forward, therefore, additional ingredients to first principles pertaining to QCD are bound to enter. It will therefore be inevitable that holes and flaws sneak in in formulating the pertinent EFT.

In this section I will treat two cases. One, normal nuclear matter and the other, highly compressed baryonic matter relevant to compact stars. What one can do here is hardly rigorous. Somewhat complicated details will be left out whenever feasible without losing the key features.

\subsection{The problem}
How the weak axial-vector coupling constant $g_A$ behaves in nuclei has a very long history.  A fundamental quantity, it has impacts on nuclear structure as well as nuclear astrophysical processes. The basic issue raised is associated with how chiral symmetry, a fundamental property of QCD, is manifested in nuclei where the presence of strongly interacting nucleons modifies  the vacuum and hence the quark condensate $\Sigma\equiv |\la\bar{q}q\ra|$ as density is increased. Although there is no rigorous QCD proofI as for high temperature  it is now widely accepted that the pion decay constant $f_\pi$ decreases, following the decrease of the condensate $\Sigma$ at increasing density, which is a fundamental property of QCD.  It seems natural then to expect that the axial coupling constant will undergo a similar {\it intrinsic} decrease in nuclear matter.

There has been a suggestion since 1970's~\cite{wilkinson} that the axial coupling constant, $g_A\approx 1.27$ determined in the matter-free space, quenches to $g_A\approx 1$ in nuclei~\cite{buck-perez}, which has invited an interpretation that it, in consistency with the dropping of $f_\pi$, also signals a precursor to chiral restoration. This has led to extensive studies accompanied by controversies in nuclear Gamow-Teller transitions, most notably giant Gamow-Teller resonances.  A review up to 1992 is found in  \cite{osterfeld}. It was deduced from $sd$-shell nuclei~\cite{brown-wildenthal} and $fp$-shell nuclei with $A\leq 49$~\cite{zuker,zuker2} and with $A=48-64$~\cite{koonin} that the effective Gamow-Teller coupling constant needed was $g_A^\ast\approx 0.8 g_A$.  In a more recent development, it was shown that with a $20\%$ quenching in $g_A$, the Gamow-Teller strengths distribution can be quantitatively reproduced by shell-model calculation in $pf$ shell for the transition $^{56}$Ni $\to ^{56}$Cu~\cite{sasano-Ni56}. It has been suggested that $g_A^\ast=1$ is  ``universally" applicable in nuclear medium. What makes this possible universality in this $g_A$ value is that it seems to insinuate partial restoration of chiral symmetry in finite nuclei.

In this Section, I revisit this long-standing problem and give an extremely simple and unambiguous answer with an effective field theory in which scale symmetry and chiral symmetry of QCD are incorporated. I predict that $g_A$ for Gamow-Teller transitions should remain {\it unaffected} by the vacuum change with density -- at least up to density $n\sim 2n_0$,  whereas in a stark contrast, it gets strongly {\it enhanced} in axial-charge (first-forbidden) transitions. This implies that if there is indeed quenching of $g_A$ in Gamow-Teller transitions in finite nuclei as it has been thought,   it cannot be the vacuum change manifested in the intrinsic chiral condensate that is responsible for it.

Nonetheless at high density $n>> n_0$, there is what is referred to as ``dilaton-limit fixed point " (DLFP) at which $g_A^\ast$ does approach 1. This comes about not because of chiral symmetry restoration as previously thought but because of the ``emergence" of scale (or conformal) symmetry in dense medium. But this density is way beyond the measurable.
\subsection{Scale-Invariant HLS Lagrangian ($s$HLS)}
My reasoning exploits the recently formulated effective field theory (EFT) Lagrangian~\cite{PKLR,PKLMR} which is valid for phenomena at low density as well as for high density appropriate for compressed baryonic matter inside such compact stars as the recently discovered $\sim 2$-solar mass neutron stars~\cite{massive-stars}.

It is constructed by implementing scale symmetry  and hidden local symmetry (HLS) to baryonic chiral Lagrangian consisting of the pseudo-scalar Nambu-Goldstone bosons,  pions ($\pi$),  and baryons. The basic assumption that underlies the construction of the effective Lagrangian, denoted $bs$HLS, with $b$ standing for baryons and $s$ standing for the scalar meson, is that there are two hidden symmetries in QCD: one,  scale symmetry broken both explicitly by the QCD trace anomaly and spontaneously with the excitation of a scalar pseudo-Nambu-Goldstone boson, i.e., ``dilaton" $\sigma$;  two, a local flavor symmetry higgsed to give massive $\rho$ and $\omega$.  Neither is visible in QCD in the matter-free vacuum, but the possibility, suggested in \cite{PKLR}, is that both can appear as ``emergent symmetries" in dense matter and control the equation of state (EoS) for highly compressed baryonic matter. I will use this same Lagrangian for calculating nuclear responses to the electro-weak current. There are no unknown parameters in the calculation.

Since the arguments are quite involved and given in great detail elsewhere~\cite{PKLR,PKLMR}, I summarize as concisely as feasible only the essential points that I need for the discussion. In addition to the nucleon and the pion, there are two additional -- massive -- degrees of freedom essential for the  $bs$HLS Lagrangian: The vector mesons $V=(\rho, \omega)$ and the scalar meson denoted $\sigma$. By now a very well-known procedure, the vectors $V$ are incorporated via hidden gauge symmetry (HLS)~\cite{HLS-PRL,HLS,HLS-loops}, which is gauge-equivalent to non-linear sigma model at low energy, that elevates the energy scale to that of the vector mass $\sim 770$ MeV.  The scalar $\sigma$ is incorporated~\cite{LMR} by using the ``conformal compensator field" $\chi$ transforming under scale transformation with scale dimension 1, $\chi=f_\chi e^{\sigma/f_\chi}$.

The underlying approach to nuclear EFT with $bs$HLS is the Landau Fermi-liquid theory anchored on Wilsonian renromalization group (RG). For this, the ``bare" parameters of the EFT Lagrangian are determined at a ``matching scale" $\Lambda_M$ from which the RG decimation is to be made for quantum theory\footnote{In what follows, ``bare" parameters put in  quotation mark will stand for these quantities defined by the matching. I will eschew using the quotation whenever feasible.}.The matching is performed with the current correlators between the EFT and QCD, the former at the tree-order and the latter in OPE. The QCD correlators contain, in addition to perturbative quantities,  nonperturbative ones, i.e., the quark condensate $\la\bar{q}q\ra$, the dilaton condensate $\la\chi\ra$, the gluon condensate $\la G_{\mu\nu}^2\ra$ and mixed condensates. The matching renders the bare parameters of the EFT Lagrangian dependent on those condensates.  {\it Since the condensates are characteristic of the vacuum, as the vacuum changes, the condensates slide with the change.} Here we are concerned with density, so those condensates must depend on density.  This density dependence, inherited from QCD, is an ``intrinsic" quantity to be distinguished from mundane density dependence coming from baryonic interactions. It will be referred to as ``intrinsic density dependence" or  IDD for short.

There are two scales to consider in determining how the IDDs enter in the EFT Lagrangian.

One is the energy scale. The (initial) energy scale is the matching scale from which the initial (or first)  RG decimation is performed.  In principle it could be the chiral scale $\Lambda_{\rm chiral}\sim 4\pi f_\pi\sim 1$ GeV. In practice it could be lower, typically just  above the vector meson mass. The scale to which the first decimation is to be made could be taken typically to be the top of the Fermi sea of the baryonic matter.

The other scale is the baryon density.  The density relevant for massive compact stars can reach up to as high as $\sim 6 n_0$. To be able to describe reliably the properties of both normal nuclear matter and  massive stars, a changeover from the known baryonic matter to a different form of matter at a density $\sim 2 n_0$ is required. In  \cite{PKLR,PKLMR}, it is a topology change from a skyrmion matter to a half-skyrmion matter. Being topological this property can be taken to be robust. In quark-model approaches, it could be the  hadron-quark continuity that encodes continuous transitions from hadrons to strongly-coupled quark matter or quarkyonic matter~\cite{fukushima}. I believe, as conjectured in \cite{PKLMR}, that the two approaches are in some sense equivalent. I will come back to this matter later.

The changeover, as I will explain later,  is not a bona-fide phase transition.  However  it impacts extremely importantly on the EoS in the formalism I am adopting, making, for instance, the nuclear symmetry energy transform from soft to hard at that density, thereby accommodating the observed $\sim 2$-solar mass stars.  Of crucial importance for the process  is that when the matter is treated in terms of topological objects, skyrmions, this changeover induces dramatic changes in the physical quantities involved in dense matter. This feature, translated to the bare parameters of $bs$HLS Lagangian, makes the IDDs differ drastically from below to above at the changeover density $n_{1/2}$, which is estimated to be around 2$n_0$~\cite{PKLR}.

It turns out that up to $n \sim n_{1/2}$, the IDD is entirely given by the dilaton condensate $\la\chi\ra$. This is because below $n_{1/2}$ the bare parameters of the Lagrangian do not depend much on the quark condensate as explained in \cite{PKLR,PKLMR}. The $\chi$ field is a chiral scalar, whereas $\bar{q}q$ is the fourth component of the chiral four vector. Therefore the dilaton condensate is not directly connected to the quark condensate, but as mentioned  below, this dilaton condensate gets locked by strong interactions to the pion decay constant which is related to the quark condensate.  While the quark condensate does not figure explicitly in the IDD at low densities,  it controls the behavior of vector-meson masses at compact-star densities, $n\gsim n_{1/2}$~\cite{PKLR}.

 \subsection{Axial Current with IDD}
 That the IDD could be entirely given by the dilaton condensate was conjectured in 1991~\cite{BR91}, and it has been confirmed to hold up to the density $n\lsim n_{1/2}$~\cite{PKLR,PKLMR}. What is new in the new development is that in the Wilsonian renormalization-group formulation of nuclear effective field theory,  this IDD-scaling is indeed all that figures up to $n_{1/2}$.  It does undergo a drastic change at $n\gsim n_{1/2}$ affecting the EoS for compact stars,  but it does not affect the axial-current problem that I address below.

 Now how the dilaton condensate takes the place of the quark condensate in the IDDs for $n\lsim n_{1/2}$ is intricate involving the role of explicit scale symmetry breaking in the spontaneous breaking. I will say a bit more on this in connection with the sound velocity of  massive stars where this issue is relevant, but let me admit that this is an issue that is not well understood even by experts. There have been extensive discussions on the matter of ``light scalars" in the effort to understand the light Higgs boson in terms of dialtonic structure for large number of flavors $N_f\sim 8$,  but little has been clarified of the matter in QCD for $N_f\sim 3$ we are concerned with. In the present case of low density, $n\lsim n_{1/2}$, however, the effect of the scale-symmetry explicit breaking, at the leading order,  turns out to be  embedded entirely in the dilaton potential, so it does not enter explicitly
% \footnote{It is of course buried implicitly in the dilaton condensate, which however is determined from the scaling $\Phi$.}
 in the axial response functions in nuclei and nuclear matter that we are interested in. This makes the calculation of the ``intrinsically modified" $g^\ast_A$ in nuclear medium extremely simple.  All we need is the part of the $bs$HLS Lagrangian, scale invariant and hidden local symmetric, that describes the coupling of the nucleon to the external axial field ${\cal A}_\mu$. Writing out explicitly the covariant derivatives involving vector fields, hidden local and external,  and keeping only the external axial vector field ${\cal A}_\mu$, and to the leading order in the explicit scale symmetry breaking, the relevant Lagrangian takes the simple form
\be
 {\cal L} &=&i\overline{N} \gamma^\mu \del_\mu N -\frac{\chi}{f_\chi}m_N \overline{N}N +g_A \overline{N}\gamma^\mu\gamma_5  N{\cal A}_{\mu}+\cdots\label{LAG}
 \ee
  Note that the kinetic energy term and the nucleon coupling to the axial field are scale-invariant by themselves and hence do not couple to the conformal compensator field. Put in the nuclear matter background, the bare parameters of the Lagrangian will pick up the medium VeV. Thus in (\ref{LAG}) the nucleon mass parameter will scale while $g_A$ will not:
 \be
 m_N^\ast/m_N=\la\chi\ra^\ast/f_\chi\equiv \Phi, \ \ g_A^\ast/g_A=1\label{Phiscaling}
 \ee
 where $f_\chi$ is the medium-free VeV $\la\chi\ra_0$ and the $\ast$ represents the medium quantities. The first relation is one of the scaling relations given in \cite{BR91}. The second is new and says that the Lorentz-invariant axial coupling constant {\it does not} scale in density. Now in medium, Lorentz invariance is spontaneously broken, which means that the space component, $g_A^{\rm s}$, could be different from the time component $g_A^{\rm t}$. Writing out the space and time components of the nuclear axial current operators, one obtains
 \be
\vec{J}_A^{\pm} (\vec{x}) &=& g^{\rm s}_A \sum_i \tau_i^{\pm} \vec{\sigma}_i \delta(\vec{x}-\vec{x}_i),\label{GT}\\
 J_{5}^{0\pm} (\vec{x})&=&- g^{\rm t}_A \sum_i\tau_i^\pm  \vec{\sigma}_i \cdot (\vec{p}_i  - \vec{k}/2) /m_N \delta(\vec{x}-\vec{x}_i)\label{axialcharge}
 % + \delta{\vec{x}-\vec{x}_i) + \frac{g^{\rm t}_A}{m_N} \sum_i \tau_i^\pm  \vec{\sigma}_i\cdot \vec{k/2}\delta{\vec{x}-\vec{x}_i) ...
 \ee
 where $\vec{p}$ is the initial momentum of the nucleon making the transition and $\vec{k}$ is the momentum carried by the axial current.  In writing (\ref{GT}) and (\ref{axialcharge}),  the nonrelativistic approximation is made for the nucleon. This approximation is valid not only near $n_0$ but also in the density regime $n\gsim n_{1/2}\sim 2n_0$. This is because the nucleon mass never decreases much after the parity-doubling sets in at $n\sim n_{1/2}$ at which $m_N^\ast\to m_0 \approx (0.6-0.9) m_N$~\cite{PKLR}.

 A simple calculation gives
\be
g_A^{\rm s}=g_A, \ \ g_A^{\rm t}=g_A/\Phi\label{main}
\ee
with $\Phi$ given by (\ref{Phiscaling}).
This is the unequivocal prediction of the  $bs$HLS with IDDs.

%\vskip 0.2cm

\subsection{Axial-Charge Transitions}
An extremely interesting process in the context of nuclear EFT, going beyond the conventional chiral perturbation theory (S$\chi$EFT), is the first forbidden (FF) process involving the axial-charge operator ${J^0_{5}}^\pm $,
\be
A(0^\pm ) \to B(0^\mp) + e +\nu, \ \ \Delta T=1.\label{AC}
\ee
In nuclear EFT,  the leading term in the chiral power counting  is the  one-body axial-charge operator (\ref{axialcharge}).
%\be
 %J_{5}^{0\pm} (\vec{x})_{\rm 1-body}=- g^{\rm t}_A \sum_i\tau_i^\pm  \vec{\sigma}_i \cdot (\vec{p}_i  - \vec{k}/2) /m_N \delta(\vec{x}-\vec{x}_i)\label{axialcharge}
 % + \delta{\vec{x}-\vec{x}_i) + \frac{g^{\rm t}_A}{m_N} \sum_i \tau_i^\pm  \vec{\sigma}_i\cdot \vec{k/2}\delta{\vec{x}-\vec{x}_i) ...
 %\ee
% where $\vec{p}$ is the initial momentum of the nucleon making the transition and $\vec{k}$ is the momentum carried by the axial current.
 From (\ref{main}), one has
 \be
 g_A^t=g_A\frac{f_\pi}{f_\pi^\ast}.
 \ee
This follows because in the approximation adopted, the ratio $\la\chi\ra^\ast/f_\chi\approx f_\pi^\ast/f_\pi$~\cite{PKLR} where  $f_\pi^\ast$ is the pion decay constant in nuclei. The ``effective"  pion decay constant is measured experimentally in deeply bound pionic systems, and it has been determined at normal nuclear matter density $n_0$~\cite{kienle-yamazaki}
 \be
 f_\pi^\ast (n_0)/f_\pi \equiv \Phi (n=n_0)\approx 0.8.\label{fpistar}
 \ee
 One has to be cautious as to whether this effective constant is {\it the} IDD appropriate as the{\it  scaling parameter} in the EFT Lagrangian we are using. This is because the experimental value is not extracted using the very EFT Lagrangian used, so it may have some corrections but it cannot be much different from the canonical value (\ref{fpistar}).\footnote{It cannot be overemphasized that this is a just {\it parameter} specific to the Lagrangian constructed, not a physical quantity, and the precise value will depend on how the EFT is defined. This point is widely misinterpreted in nuclear physics community, which has led to some absurd statements on the idea of \cite{BR91}.}

\begin{figure}[h]
%\vskip -5.cm
\begin{center}
\includegraphics[width=6cm]{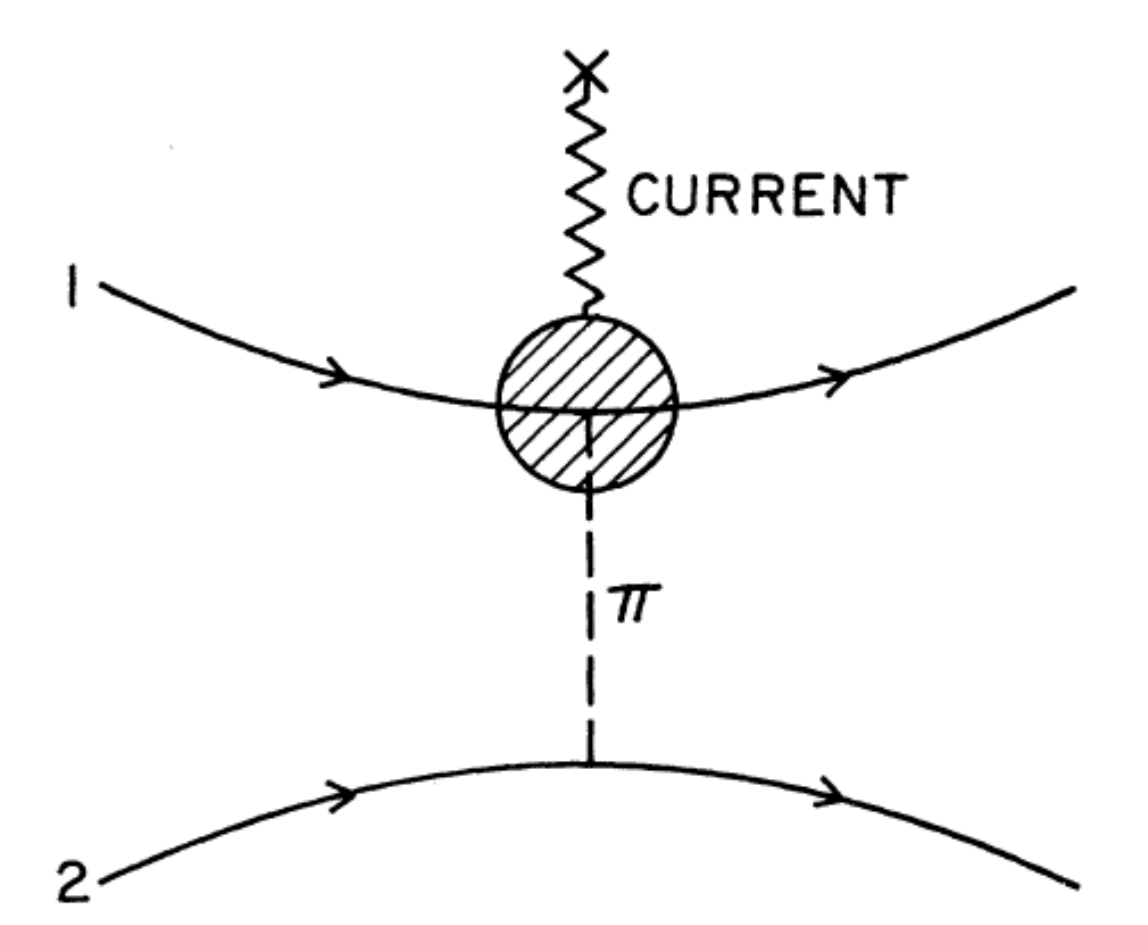}
%\centerline{\epsfig{file=cc.pdf, width=9cm}}
\end{center}
\caption{Two-body exchange current. The upper vertex involves two soft pions for the axial charge transition.}\label{2-body}
\end{figure}

 To the next order in the chiral counting, two-body currents contribute. The most important one is the pion-exchange term depicted in Fig.~\ref{2-body}.
Consider the upper vertex. One can take the time component of the axial field as soft in the first-forbidden beta decay that involves long wave-length probe.  With the exchanged pion also soft,  the vertex involves two soft pions controlled by the current algebra.  It is $O(1)$ in the chiral counting and hence will make a huge  contribution. This was recognized already in 1978. In terms of chiral Lagrangian, this is automatically encoded at the leading chiral order. It is easy to work out the two-body charge operator~\cite{KDR}
\be
J_{5}^{0\pm} (\vec{x})_{\rm 2-body}= g_A^t \Phi^{-1} \frac{m_\pi^2}{8\pi f_\pi^2}\sum_{1<j}(\tau_i\times\tau_j)^\pm [\vec{\sigma}_i\cdot\hat{r}\delta(\vec{x} -\vec{x}_j) +(i\leftrightarrow j)]{\cal Y}(r)\label{two-body-cahrege}
\ee
with
\be
{\cal Y}=\frac{e^{-m_\pi r}}{m_\pi r}\Big(1+\frac{1}{m_\pi r}\Big).\nonumber
\ee
Given accurate wavefunctions, the transition matrix elements of both the one-body and two-operators can be calculated extremely accurately. The ratio $R$
\be
\frac{R}{\Phi}=\frac{\la B|J_{5}^{0\pm} (\vec{x})_{\rm 2-body}|A\ra}{\la B|J_{5}^{0\pm} (\vec{x})_{\rm 1-body}|A\ra}\label{Ratio}
\ee
is found to be highly insensitive to the wave functions and hence to the density of the system.   One gets
\be
R=0.5\pm 0.1
\ee
ranging in mass number from $A=12$ to $A=280$. This can be understood by the fact that the two-body operator is long-ranged given by the soft-pion exchange, totally controlled by chiral symmetry. Note that the ratio of the 2-body over the one-body matrix elements (\ref{Ratio}) is BIG, $\sim 0.6$ for the nuclei involved.\footnote{This is the largest exchange current correction in nuclei I know of outside of anomalous cases where accidental cancellations are involved.}

The effective axial-charge operator is then given by replacing $g^{\rm t}_A$ in (\ref{axialcharge}) by the effective one
\be
{g_A}^{\rm eff}=g_A^t (1+\frac{R}{\Phi}).\label{A0}
\ee
Precise experimental data are available for the FF  transitions in the Pb region $A=205-212$~\cite{warburton}.  The analysis in \cite{warburton} is given in terms of the medium-renormalization of the free-space axial-coupling constant $g_A$
\be
\epsilon={g_A}^{\rm eff}/g_A.\label{epsilon}
\ee
The prediction for $n=n_0$ given by  (\ref{epsilon}) with (\ref{A0})  comes out to be~\cite{KR}
\be
\epsilon(n_0)_{theory}=2.0\pm0.2
\ee
to be compared with the experiment in the Pb region~\cite{warburton}
\be
\epsilon_{exp}=2.01\pm0.05.
\ee
This leads to a large enhancement in the FF transitions.

This huge enhancement  could have an important impact in astrophysical processes.  For instance it was observed~\cite{honmaetal} that in $N=126$ isotones, the first-forbidden transition -- which is relevant to astrophysics -- affects appreciably the half-lives for larger $Z$: For  $Z=72$, the half-life is quenched by a factor of $\sim 5$ by the  FF contribution. This calculation however does not take into account the $\epsilon$ factor. If the FF process is dominated by the single-particle axial charge operator in this process, a back-of-envelope estimate shows that the soft-pion-exchange contribution could make the half-life an order of magnitude shorter than the estimate given in  \cite{honmaetal}.
It would be  interesting to re-do the calculation  with this enhancement effect duly taken into account and see how it affects the astrophysical process in question.

Similar enhancements are found in precision experiments in the lighter nuclei $A=12-15$~\cite{O16,B12}. This confirms the universality of the effect. In this mass range, an extremely accurate calculation in full consistency with chiral perturbation theory should be feasible. Higher-order corrections to the two-body axial-charge operator, e.g., multi-body currents, are strongly suppressed making the calculation free of uncertainty, thereby offering a powerful proof to the FT.  If the process probed short-range interactions, then short-range multi-body currents could come in importantly, but not in long-ranged processes considered here. This could be considered also as a pristine evidence for the soft-pion dominance in certain nuclear processes unhampered by higher-order intrusion.

%The role played by soft pions in the process. Considering the soft probe, $A_0$, as pionic, one can think of the upper vertex of Fig.~\ref{2-body} as involving two soft pions. This is the core of the current algebras of 1960, which is now fully captured in nonlinear sigma model with derivative coupling, and hence in currently successful effective quantum field theory.
%This can be considered as a striking case of Weinberg's ``folk theorem" on effective quantum field theory ``proven" in nuclear physics~\cite{weinberg}.
%In fact this matter, dating from 1970's, illustrates that ``what nuclear physicists have been doing all along is the correct first step in a consistent approximate scheme."

What's discussed above raises the issue on the role of NG bosons in nuclear physics. When the energy scale probed in nuclear processes is much less than the pion mass $\sim140$ MeV, the pion can also be integrated out in the spirit of the FT and one obtains ``pionless effective field theory ($\not\pi$EFT)" for nuclear physics, which is eminently a respectable effective field theory.   But  how do the soft-pion theorems, encoded in chiral Lagrangians, manifest in $\not\pi$EFT? It is not at all obvious that a term of this nature is captured in the pionless treatment.
With no pions present,  there is no explicit footprint of chiral symmetry for the spontaneous breaking of chiral symmetry. Yet the theory seems to work fairly successfully in various low-energy processes involving light nuclei, e.g., the $pp$ fusion, the thermal $np$ capture etc.  The  solar proton fusion process is dominated by the Gamow-Teller operator, so it could very well be captured. But what about the double-soft process that makes such a huge effect in the axial-charge transitions?  Is it hidden in the pionless theory?

\subsection{Gamow-Teller Transitions}
The Gamow-Teller (GT) transition in nuclei is given by the space component of the axial current (\ref{GT}).  Unlike the time component treated above,  this transition receives no soft-pion contribution and the matrix element is predicted to be unaffected by the IDD as given in (\ref{main}) at least up to the density $n_{1/2}\sim 2n_0$. This means that the space component of $g_A$  should remain unmodified by density in contrast to  the pion decay constant $f_\pi$ which drops at increasing density, signaling partial restoration of chiral symmetry.  What this means in nuclear GT transitions can be seen  in the so-called GT sum rule. Consider the operator identity
\be
(S^- - S^+) = 3(N - Z)\label{SR}
\ee
where $S^-$  and $S^+$ are the transition strengths, corresponding to the square of the matrix elements of the operator $\sum_i \tau_i^{\pm} \vec{\sigma}_i$
summed over all final states. Now suppose one measures the Gamow-Teller strengths, corresponding to the operator Eq.~(\ref{GT}), in experiments.  Then the prediction (\ref{main}) says that  if two-body and higher multi-body exchange currents were ignored, which is justified in S$\chi$EFT for allowed Gamow-Teller transitions, one would expect that the GT sum rule with the strengths multiplied by $(g_A^\ast)^2$ should be satisfied by $g_A^\ast\approx g_A$, the free-space $g_A\approx 1.27$. In other words, there will be no quenching. What does this mean in giant Gamow-Teller resonance processes in nuclei?

At first sight, such a GT sum-rule result seems to be at odds with the QCD sum-rule calculation~\cite{drukarev} which finds  $g^\ast_A$ is quenched to 1 as the quark condensate decreases. The QCD sum-rule result seems to be consistent with the observations made in light nuclei~\cite{buck-perez}, namely that the Gamow-Teller transition in light nuclei can be described in simple shell-model treatments with the ``intrinsic" $g_A^\ast = 1$.  On the other hand, the theory developed in \cite{PKLMR}, as discussed above, suggests that the GT strength should be governed by (\ref{Phiscaling}) and hence unquenched $g_A$ in nuclei.  It is only at what is called ``dilaton-limit fixed point" at high density, $n > n_{1/2}\sim 2n_0$, where scale symmetry gets unhidden,  that $g_A^\ast$ should approach 1.  That the GT sum rule might be saturated with very little quenching has been discussed~\cite{sakai-suzuki,sakai-yako}.

That the effective $g_A^\ast$ in Gamow-Teller transitions in light nuclei described in a simple shell model  is quenched from the free-space value has been argued to be explainable by ``nuclear core polarizations"~\cite{osterfeld} that are corrections to the simple shell-model wave functions constructed within a limited shell. In this description, it is the higher energy-scale  multi-particle-multi-hole configurations  (e.g., high-order core polarizations) ``integrated out" (in the language of effective field theory) that result in the renormalization of the axial coupling constant. If one were to do a full shell model calculation in the space of nucleon configurations, this quenching would not be needed.  This should be verifiable  by realistic ``no core" shell model calculations that are being actively developed.  Should this be confirmed, then why the effective $g_A^\ast$ in nuclei in single-shell (e.g., $fp$, $sd$ etc)  model (analog to mean-field theory) is {\it universally}  very near 1 as observed would then remain mysterious.

From the point of view of EFT \`a la FT, that the Gamow-Teller sum rule may be entirely or mostly satisfied by $g_A^\ast=g_A$, i.e., unquenched, is actually unnatural for the following reason. Although the IDD -- effective from the matching scale -- does not affect the $g_A^s$, one still has to consider what other degrees of freedom intervening at scales lower than the matching scale $\sim 1$ GeV can contribute in the renormalization decimations involved in the EFT Lagrangian. The first energy scale one encounters in the Gamow-Teller channel as one makes the decimation from the matching scale  is the $\Delta$-hole excitation energy $E_{\Delta-h}$ lying $\sim 300$ MeV above the Fermi sea.  This channel is strongly coupled to nucleon-hole states while other resonance channels can be ignored, so there is no reason why this $\Delta$-hole effect cannot contribute significantly to the Gamow-Teller response functions in the decimation down to $E_{\Delta-h}$. Since the weak current acts only once, this effect can be included in the modification of the Gamow-Teller coupling constant $g_A^s$ to $g_A^\ast\neq g_A$. From that scale it will then be purely the correlations involving nucleons only that should figure, therefore the duly renormalized  constant $g_A^\ast$ should be effective in full, {\it no core}, shell-model calculations. This renormalized constant was worked out a long time ago, which could be phrased in terms of Landau-Migdal's ${g_0^\prime}$ parameter in the $\Delta$-N channel~\cite{mr73,ohta-wakamatsu,oset-rho,maruyama}  when treated in Landau's Fermi liquid theory in the space of $N$ and $\Delta$. This can also be phrased in terms of the Ericson-Ericson-Lorentz-Lorenz (EELL) effect in pion-nuclear interactions~\cite{EELL,delorme}.  Since  Landau-Migdal theory works excellently in the nucleon space, it seems reasonable to expand the space to both spin-1/2 and spin-3/2 spaces and treat the interactions in three (coupled) channels, NN, $\Delta$N and $\Delta\Delta$. In this generalized Fermi-liquid theory, the Gamow-Teller process will then involve the Landau-Migdal's quasiparticle interactions $g_0^\prime (\sigma_1\cdot\sigma_2)( \tau_1\cdot\tau_2)$ in the three channels.  If one takes the universal value for all three channels, $g_0^\prime|_{NN}=g_0^\prime|_{\Delta N}=g_0^\prime|_{\Delta\Delta}$, then from the $g_0^\prime|_{NN}$ determined  from pion-nuclear interactions, one predicts that $g^\ast_A$ for Gamow-Teller transitions in nuclear matter is renormalized to $g_A^\ast\approx 1$.  However analysis of experiments on giant Gamow-Teller resonances in terms of the Landau-Migdal parameters~\cite{sakai-suzuki,sakai-yako} indicates that  $g_0^\prime|_{\Delta N}$ could be considerably smaller than $g_{NN}^\prime$, violating the ``universality." Whether and how this can be related to the QCD sum-rule result~\cite{drukarev}  is not clear. What is clear, however, is that this effect is {\it not} the IDD inherited from QCD at the matching scale in the $bs$HLS formulation. Therefore it cannot be taken as a precursor to chiral restoration as the pion decay constant going to zero is.  It is just a renormalization due to ``mundane" correlations in the hadronic sector generalized to the $\Delta$-N space, not directly tied to the quark condensate, the order parameter of  chiral symmetry.

It may be possible  to argue away the presence of  the $\Delta$ degree of freedom for the $g_A$ problem. Given that nucleon-hole states can be strongly excited by the tensor force to the excitation energies $\sim (200-300)$ MeV comparable to $\Delta$-hole states, taking into account the $g_0^\prime$ effect in the $\Delta$-hole channel may be {\it simulated} by multi-particle-multi-hole states of the comparable energy-scale within the nucleon space without explicit $\Delta$-hole states. Thus it may be the Gamow-Teller sum rule  (limited to the nucleon space)  could be saturated with nearly unquenched $g_A$ in a full scale ``no core" calculations. It may be that by going higher in energy scale with multi-particle-multi-hole states, $g_0^\prime|_{\Delta N} << g_0^\prime|_{NN}$ could be accommodated. However this does not  invalidate the result that exploits the $\Delta$-hole excitations~\cite{BR-GT,bohr-mottelson}. It is just that with the strong spin-isospin excitations involved, the nucleon-hole space  and $\Delta$-hole space may not be sharply delineated.  From the point of view of quark models, there is no qualitative difference between nucleon-nuclear interactions and $\Delta$-nuclear interactions. There is an analogy to this dichotomy in the Carbon-14 dating -- discussed below -- where part of three-body forces and ``Brown-Rho scaling" play the same role.  In this connection, the argument made by Gerry Brown in his unpublished note~\cite{GEB} is highly pertinent: that ``the $g_0^\prime$ interaction is by far the most important interaction between nucleon quasiparticles and together with the Brown-Rho scaling, runs the show and make all forces equal."
\subsection{Postdicting the Hoyle state and the  Carbon-14 dating}
\subsubsection{The Holye state}
Among various applications of the FT to nuclear physics,  the calculation of the Hoyle state is interesting in two ways. The Hoyle state plays a key role in the helium burning of heavy stars and in the production of carbon and other elements.

Now with the great development of the numerical techniques mentioned above as {\it ab initio}, the many-body problem for the mass number $\gsim 12$ could be addressed,  given accurate potentials. Indeed this state has been calculated with Monte Carlo lattice simulation using the potential calculated to ${\cal O} (Q^3)$ in S$\chi$EFT, successfully obtaining the Hoyle state as well as low-lying spin-2 rotational bands~\cite{epelbaumetal}. It seems plausible that highly sophisticated phenomenological potentials could do equally well in the sense of MEEFT. I think this result nicely  illustrates the working of both the FT and its Corollary.

That the FT could work in practice in various different ways is aptly illustrated in the Skyrme model description. The Skyrme model -- with pion field only -- is shown to correctly describe the Hoyle state and its rotational band as well as the ground-state rotational band~\cite{lau-manton}. The ground state is interpreted as an equilateral triangle of $B=4$ ($B$ being the topological winding number) skyrmions, i.e., alpha particles, and the Hoyle state is a linear chain. As discussed below, other degrees of freedom, such as vector mesons and scalar dilaton,  figure in the skyrmion dynamics for dense matter. For the Hoyle state and its ground state properties, however,  what matters is the topology that is carried by the pion field. Thus topology must play an essential role in giving the correct structure. There is no translation between this skyrmion description and the Monte Carlo lattice calculation of \cite{epelbaumetal}, so it is difficult to see how the FT works here. There seems also to be some basic difference in the structure of the Hoyle state: The {\it ab initio} lattice calculation seems to favor an obstruce triangular structure in contrast to the linear chain of alpha particles in the skyrmion picture. In any case one intriguing possibility is that as I will develop below, one could resort to the Cheshire Cat mechanism that trades in topology of the skyrmion structure for the quark-gluon structure of QCD in an EFT. Ultimately one could invoke a similar mechanism to relate the two pictures of the  Hoyle state. In fact this ``trading" of topology for certain properties of field theory will be exploited below in formulating the $V_{lowk}$RG formalism to access compact star matter.
\subsubsection{The carbon-14 dating}
The beta decay transition from the initial state ($J^\pi,T)= (0^+,1)$ of $^{14}$C to the final state  ($J^\pi,T)= (1^+,0)$ of $^{14}$N is singularly retarded with the half-life of 5730 years despite that  it involves allowed Gamow-Teller transition. Explaining this anomalous behavior illustrates the extreme subtlety in accessing dense matter in effective field theory. It was shown~\cite{holt-c14} that the wave functions involved in the transition probe the density of $\sim 0.75n_0$ and it is at this density that the Gamow-Teller matrix element nearly vanishes. The cancellation in the matrix element is extremely sensitive to the decrease in the nuclear tensor force as density increases towards $n=n_{1/2}\sim 2n_0$ (see Fig.~\ref{tensorF} below). This behavior will turn out to be a key feature in going to compact stars described below. Although $^{14}$C is a light nucleus,   it is at density near $n_0$ that the overlap of the wave functions peaks. The interplay between the cancellation in the Gamow-Teller matrix element and the splitting between the $1_1^+$ and $0_1^+$ in $^{14}$N provides a strong support to the validity of this interpretation. In \cite{holt-c14}, the density-scaling of the tensor force is simulated with an effective density dependence in the parameters of the exchanged mesons in the two-body forces, referred by the authors to as ``Brown-Rho scaling."  It is important to note that this contains not just the IDD defined above but also ``induced" effect that I will explain.

It has been shown that with the introduction of chiral three-body forces, the suppressed rate can be equally well explained~\cite{holt-kaiser-weise:c14}. What is significant in this treatment is that the principal mechanism is provided by the short-range three-body force. In terms of the $bs$HLS Lagrangian, such short-range interactions can be generated by the vector-meson exchanges, in particular, one and two $\omega$ exchanges.  In the $V_{lowk}$RG approach with $bs$HLS of \cite{holt-c14}, the decimation is done from a cutoff slightly below the vector-meson mass. The effect of three-body interactions exchanging massive mesons that give rise to the contact interactions are to be integrated out and hence will  be effectively lodged in the parameter for the tensor forces as an ``induced density-depence" correction to the IDD discussed above. Thus the effective IDDs figuring in \cite{holt-c14} contain the effect of contact three-body interactions.\cite{holtetal}. It does not contain, however, long-ranged three-body force effects. Such long-range three-body effects will have to be included in the $V_{lowk}$ approach as a higher-order correction.

Suppose one includes three-nucleon interactions in the $V_{lowk}$ calculation. The three-body interactions exchanging pions, vector mesons and the dilaton, would then consist of long-ranged interactions involving pion exchanges and short-range ones involving vectors with appropriate IDD's incorporated. It is likely that the short-range three body force generated in this approach with no parameters will have its strength much reduced from what is needed in \cite{holt-kaiser-weise:c14} without IDD's which is just an unconstrained parameter.

\subsection{Venturing into Compact-Star Matter}
Beyond the  density $\sim 2n_0$, there is no help from experiments and lattice QCD. Thus going into the regime where compact-star physics takes place is an adventure. How to apply the FT in this regime is therefore an effort anchored on no solid ground.  I describe here what has been done in going to compact stars based on the $bs$HLS Lagrangian. Here again Korean nuclear theorists enter, making crucial contributions. The key role was, and continues to be,  played by the young Korean graduate student Won-Gi Paeng. What is described below is based on his work, first as his PhD thesis and then his on-going work  at RAON in the Institute for Basic Science (IBS) in Korea.

Since $bs$HLS is defined to be applicable below the the chiral scale, it seems reasonable that the dense matter relevant in compact stars with density reaching $n\sim (5-6)n_0$ can be accessed {\it entirely} without invoking other degrees of freedom than what's in the Lagrangian. For this, we opt to resort to information available from topological structure of baryons present in the Lagrangian, i.e., skyrmions present in scale-invariant HLS Lagrangian, at higher density $n \gsim 2n _0$. The basic premise is that quark-gluon dynamics at low energy can be traded in for topology, representing quark-hadron continuity. This derives from that in the large $N_c$ limit, skyrmions are the baryons in QCD~\cite{witten-largeN}. This trade-in is modeled in the Cheshire-Cat phenomenon~\cite{CC,CC2,cc-mr} where it is suggested that where and how the trade-in is done is a gauge degree of freedom.  This picture has been extended to holographic QCD coming from gravity-gauge duality in string theory~\cite{cc-holography}.  The strategy is to extract from the skyrmion model certain topological properties that are robust. This comes about because, as stressed,  topology is carried uniquely by the pion and is not affected by other degrees of freedom involved in the dynamics. These robust properties are translated into the bare parameters sliding with density of the $bs$HLS Lagrangian.  The changeover takes place at the density denoted as $n_{1/2}$ at which skyrmions fractionalize into half-skyrmions.
\subsubsection{Skyrmion-half-skyrmion transition}
Skyrmions put on crystal lattice to simulate density effect are found to undergo a transition from skyrmions to half-skyrmions at a density $n=n_{1/2}$~\cite{park-vento}.\footnote{While the existence of the skyrmion fractionalization is robust depending only on the presence of pions in strong correlations, the precise value of $n_{1/2}$ does depend on what degrees of freedom are included. With the vector and scalar mesons included as in $s$HLS, this transition comes about at $\sim 2n_0$. This is an appropriate density where there could be a smooth transition from baryons to quarks in the sense of quark-hadron continuity.}  Another graduate student at Seoul National University, Hee-Jung Lee,  contributed to,  and wrote his thesis on,  this development~\cite{hjleeetal,hjleeetal2}.  To calculate quantum effects beyond the mean field in the field theory to treat many-nucleon correlations, an efficient and powerful way is to resort to the renormalization-group (RG) flow analysis \`a la Wilson in the presence of Fermi sea~\cite{shankar,polchinski,froehlich}. The density effects due to both the sliding vacuum implemented in IDDs and nuclear correlations are captured in the RG approach. There are various different approaches doing RG in nuclear theory. The approach most versatile with the $bs$HLS Lagrangian is the $V_{lowk}$ formalism~\cite{Vlowk}. In this approach, one does the double decimations, the first decimation defining $V_{lowk}$ with the sliding vacuum effect and higher-order correlations taken into account in the second decimation. The advantage of this RG approach is that it is most faithful to the FT theorem and is close to Landau Fermi-liquid fixed point theory~\cite{stonybrook-fermi-liquid}. This procedure as applied to the compact-star physics is detailed in \cite{dongetal,PKLR,PKLMR}.

There are two important  effects of topology change at $n\gsim n_{1/2}$ that can be imported into the RG with $bs$HLS. The first is that what is equivalent to the quark condensate in QCD variables $\Sigma\equiv \la\bar{q}q\ra$ where $q$ is the light-quark field $u$ and $d$ (and sometimes $s$ for three flavors) vanishes when averaged over the unit cell of the crystal lattice (with the FCC favored energetically), which will be denoted as $\overline{\Sigma}=0$. The condensate is not zero locally.  It's only the average over the cell that vanishes. It supports chiral density wave. Though globally zero, this does not imply that chiral symmetry is restored since there is pionic excitation in the system. This means that $\Sigma$ is not the order parameter for chiral symmetry. The pion decay constant is found to be non-zero~\cite{park-vento}. Thus the topology change does not involve phase transition in the sense of Ginzburg-Landau-Wilson (GLW) paradigm. I will therefore eschew referring to it as a phase transition and call it  ``topology change."

\begin{figure}[ht]
\centering
\includegraphics[width=9cm]{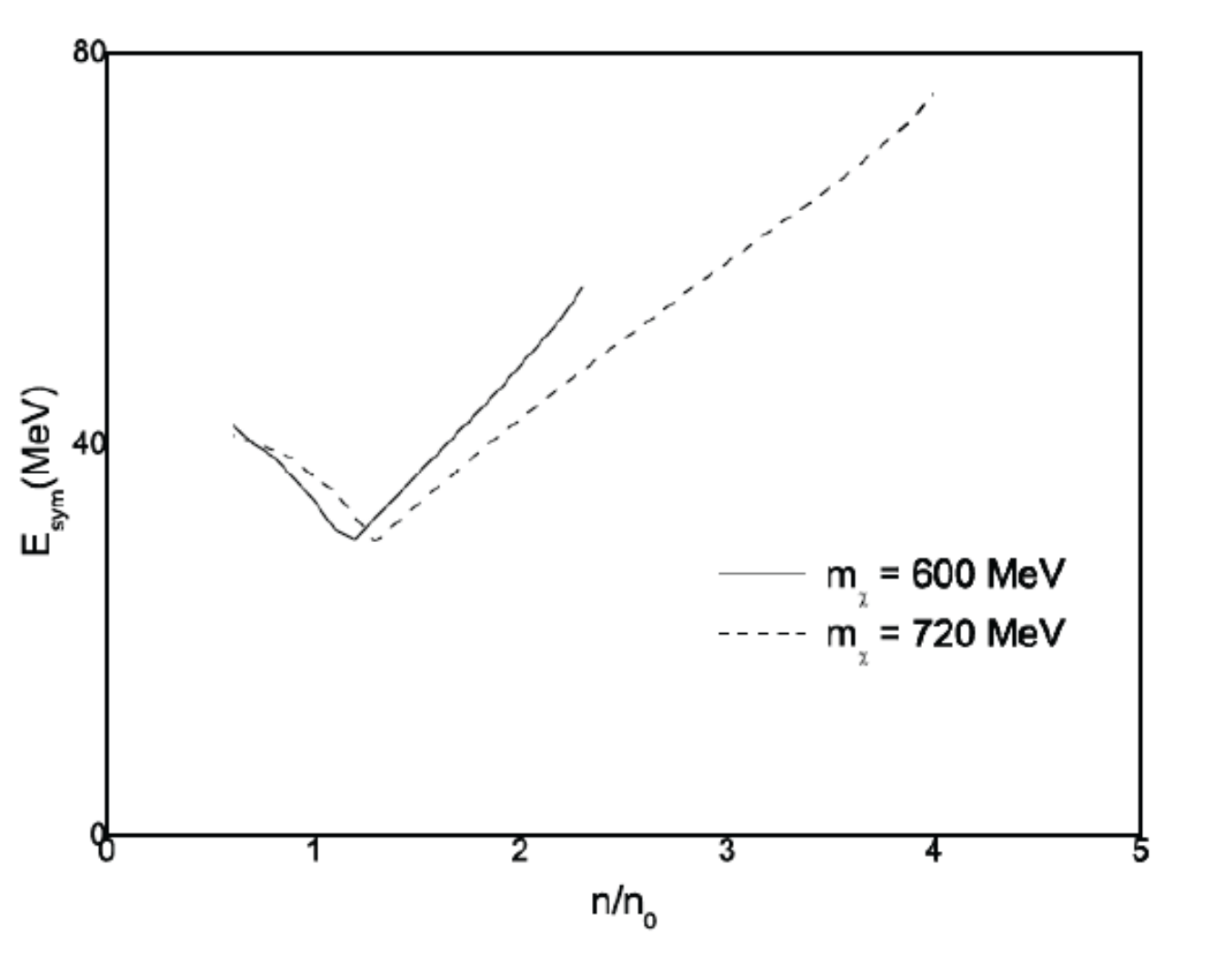}
\caption{Symmetry energy as a function of density calculated  in skyrmion crystal model. The cusp is located at $n_{1/2}$. The lower
density part  is not shown because the crystal approach and the collective quantization method used are not applicable in
that region. The location of the changeover density depends a bit on the scalar dilaton mass, which is shown for two different masses.}
\label{cusp}
\end{figure}

The second observation is in the nuclear symmetry energy $E_{sym}$ defined in the energy per particle $E(n,\alpha)$ as
\be
E(n,\alpha)=E_0 (n,\alpha=0)+\alpha^2 E_{sym} + \cdots
\ee
where $\alpha=(N-Z)/(N+Z)$ with $N(Z)$ standing for the neutron(proton) number in the nuclear system of mass $A=N+Z$. The symmetry energy $E_{sym}$ represents the energy change due to the neutron excess $\alpha > 0$ and plays an extremely important role in asymmetric nuclear systems, specially in neutron stars. $E_0$ is the energy per particle for the familiar symmetric nuclear matter. In the skyrmion description of many-nucleon systems,  $E_{sym}$ is given by  $1/ (8{\cal I}_\tau)$~\cite{LPR-cusp}  where ${\cal I}_\tau$ is the isospin moment of inertia coming from  the rotational quantization of the soliton, so it is an effect of ${\cal O} (N_c^{-1})$ to be compared with the leading term of $E_0\sim {\cal O} (N_c)$. The behavior of ${\cal I}_\tau$ as a function of density is controlled by the soliton structure, hence principally on the hedgehog pion field, and is not sensitive to other degrees of freedom. In fact it is characterized by the cusp structure as shown in Fig.~\ref{cusp}.

\subsubsection{From the cusp to the nuclear tensor force}
The discovery of the cusp in the symmetry energy, made principally by Byung-Yoon Park in his weekly visits to the newly established Korean Institute for Advances Studies (KIAS) in early 2000s, is a crucial element in the development that follows. Whether the whole idea is right or wrong hangs principally on this observation and the interpretation thereof in constructing the effective field theory.\footnote{It is perhaps worth recounting here an event that took place with this development. When the cusp was discovered, Byung-Yoon and his collaborators, Hyun Kyu Lee and the author of this article, excited about the totally unexpected result and its implication on compact stars, submitted a short note to Physical Review Letters. It was rejected by the referees. Disagreeing with the referees' assessment and verdict on its suitability in PRL, we submitted arguments rebutting their conclusion. Since the referees and the authors could not come to an agreement,  the Divisional Associate Editor at the time (circa 2010)  was recruited to intervene. The DAE formally rejected the paper on the ground -- to paraphrase -- that ``all the nuclear theorists abandoned the skyrmion model, so this paper cannot be accepted for publication in PRL." It is true that with daunting mathematical difficulties and most of the mathematically astute theorists turned to string theory, there was little progress in nuclear circle and most of the frustrated nuclear theorists dropped the model and moved over to S$\chi$EFT, which was becoming fashionable. An irony however was that at the time the cusp was discovered, skyrmions were having  a series of remarkable breakthroughs, though at lower dimensions,  in condensed matter physics. In fact,  the key idea in that article (submitted to  PRL) was inspired by certain phase transitions associated with  ``deconfined quantum critical phenomenon," chiral superconductivity etc. involving half-skyrmions or merons and monopoles. (This development in condensed matter and also in string theory for holographic baryons is found in the recent volume~\cite{multifacet}.)  It is that idea that inspired what's presented in this article. The power of topological concepts encoded in the skyrmion is more widely recognized and duly appreciated in all other areas of physics than in nuclear physics, precisely the area for which Skyrme came up with the original, revolutionary idea.  Currently, fascinating discoveries in topological phase transitions, including possibly quantum computation,  are witnessing  the wealth of marvels buried in the idea. Our response to the DAE at the time was that what's visible for skyrmions as a whole, and in particular in nuclear physics, is just ``the tip of a giant iceberg," much to be encouraged to explore, rather than to be suppressed. The recent development in compact-star physics, I claim, is a support for the idea presented in that paper. Some of the recent developments in this line of research in nuclear and astro physics are collected in a monograph volume~\cite{MaRho}.}

Taking that this cusp is a robust feature coming from topology, and generic in the dynamics, the next step is to reproduce this feature by fixing the bare parameters of the $bs$HLS Lagrangian sliding with density. In order to do so, let us see how to understand it from the point of view of an EFT with the given Lagrangian. In the standard nuclear physics approach (SNPA),  the symmetry energy is known to be controlled predominantly by the nuclear tensor force $V^T$. The tensor force acting on the ground state excites strongly the states lying at $\sim 200$ MeV above the ground state.   Using the closure approximation, which is the first approximation, the symmetry energy is then given by~\cite{brown-machleidt}
\be
E_{sym}\approx c\frac{\la |V^T|^2\ra}{200\ {\rm MeV}}\label{closure}
\ee
where $c$ is a dimensionless constant independent of density. To reproduce the cusp with this formula, the tensor potential needs to decrease as $n_{1/2}$ is approached from below and then increase after $n_{1/2}$. Let us therefore look at the tensor force effective in medium. In the $V_{lowk}$ formalism with $bs$HLS with IDDs, the tensor force is given by the sum of the pionic and $\rho$ exchange contributions. Taking the nucleon to be nonrelativistic in the domain of density relevant to compact stars, which is a good approximation since the nucleon mass remains ${\cal O} (m_0)\sim m_N$, the pion and $\rho$ tensors are of the form

\be
V^T_M \left( r \right) &=& S_M \frac{f_{NM}^{\ast\,2}}{4\pi} \tau_1\, \tau_2\, S_{12}{{\cal I}(m^\ast_{M} r)} \label{TensorM}\\
{{\cal I}(m^\ast_M r)}&\equiv& m_M^\ast
\left( \left[ \frac{1}{(m_M^\ast r)^3} + \frac{1}{(m_M^\ast r)^2} + \frac{1}{3m_M^\ast r} \right] e^{-m_M^\ast r} \right)\,,\label{radial}
\ee where $M=\pi,\,\rho$, $S_{\rho(\pi)} = +1(-1)$ and
\begin{equation}
S_{12} = 3 \frac{\left(\vec{\sigma}_1 \cdot \vec{r} \,\right)\left(\vec{\sigma}_2 \cdot \vec{r} \,\right) }{r^2} - \vec{\sigma}_1 \cdot \vec{\sigma}_2
\end{equation} with the Pauli matrices $\tau^i$ and $\sigma^i$ for the isospin and spin of the nucleons with $i = 1,2,3$.
A qualitatively important quantity is the ratio $f_{NM}^\ast/f_{NM}$
\begin{equation}
{\cal R}_M\equiv \frac{f_{NM}^\ast}{f_{NM}} =  \frac{g_{M NN}^\ast}{g_{M NN}} \frac{m_N}{m_N^\ast} \frac{m_M^\ast}{m_M}
\end{equation}
where $g_{MNN}$ are the effective meson-nucleon couplings. What is the most notable  in (\ref{TensorM}) is that given the same radial dependence, the two forces (through the pion and $\rho$ meson exchanges) come with an opposite sign.

First, we discuss the d(ensity)-scalings of the two tensor forces in medium given by IDDs and predict how the net tensor force depends on density. For the $\pi$ tensor force, we have
\be
{\cal R}_\pi =  \frac{g_{\pi NN}^\ast}{g_{\pi NN}} \frac{m_N}{m_N^\ast} \frac{m_\pi^\ast}{m_\pi} \approx  \frac{m_\pi^\ast}{m_\pi}.
\ee
 Thus the $\pi$-tensor force principally depends only on the d-scaling of the pion mass.

\begin{figure}[ht]
\begin{center}
\includegraphics[width=10.0cm]{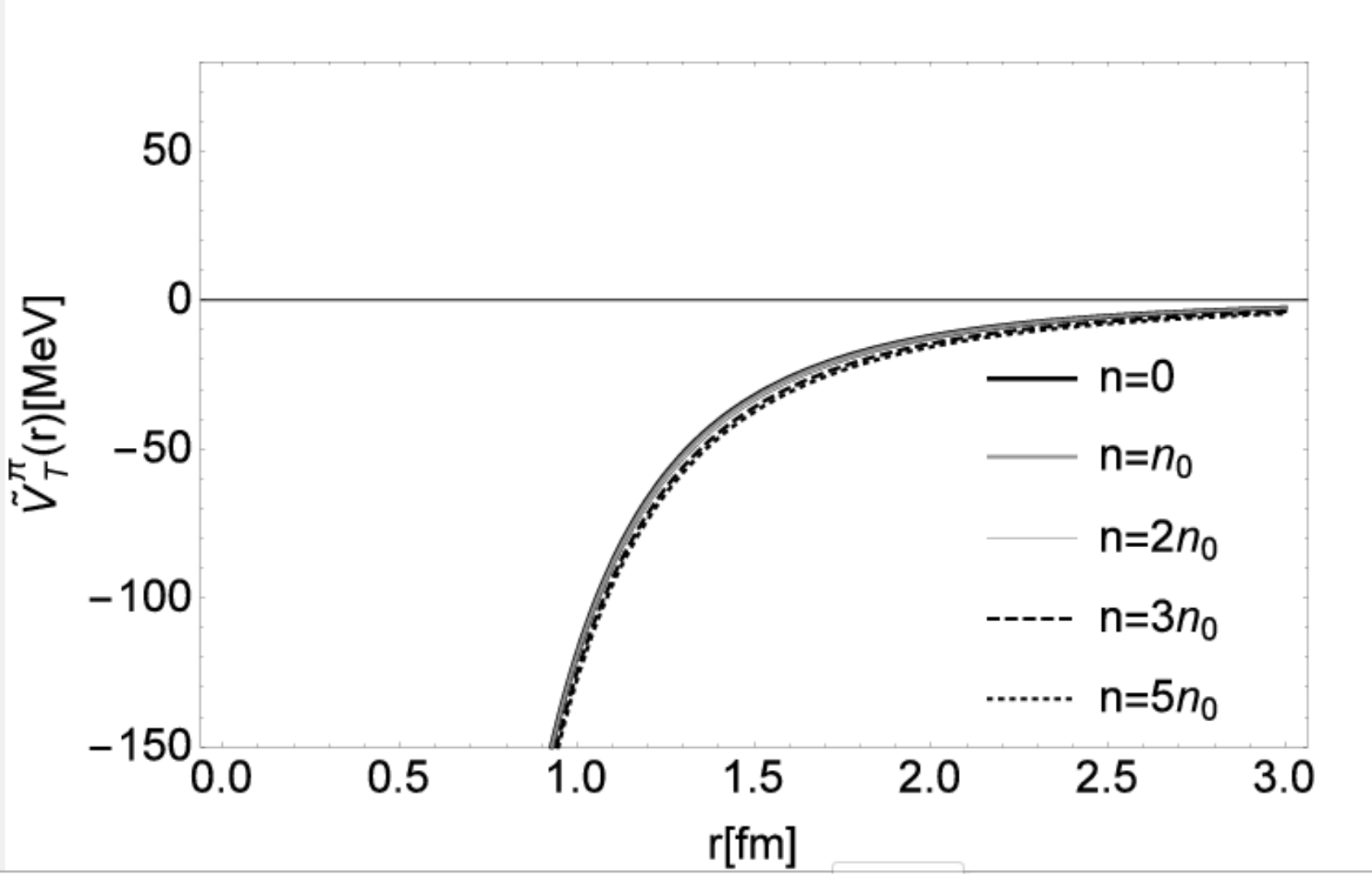}
\caption{ $\tilde{V}^\pi_T\left(r\right) \equiv V^\pi_T\left(r\right) \left( \tau_1\, \tau_2\,S_{12}\right)^{-1}$ with $n_{1/2} = 2n_0$. }
\label{pi_tensor}
\end{center}
\end{figure}

 The behavior of the in-medium pion mass is a very subtle matter. This is because the pion mass is negligible on the scale of chiral symmetry and  depends on the anomalous dimension of the quark mass term, which is not known. Whatever the case is, it is fair to take it non-scaling in the density regime considered.  A careful analysis in \cite{PKLMR} taking into account various mechanisms shows that this is confirmed. The result obtained in \cite{PKLMR} is summarized in Fig.~\ref{pi_tensor}.

Now look at the $\rho$ tensor.  The crucial quantity is the raitio
\be
{\cal R}_\rho &=&\frac{g_{\rho NN}^\ast}{g_{\rho NN}} \frac{m_N}{m_N^\ast} \frac{m_\rho^\ast}{m_\rho}.
\ee
Both theoretically and experimentally, up to nuclear matter density, the ratio is expected to be density-independent,
\be
{\cal R}\approx 1.
\ee
While the pion mass does not change with density, the $\rho$ mass scales as $m_\rho^\ast/m_\rho\approx \la\chi\ra^\ast/\la\chi\ra\approx f_\pi^\ast/f_\pi$. Thus with the opposite signs, the two tensor forces will start cancelling each other, with the $\rho$ tensor becoming greater in strength, as density goes up. It is reasonable to assume that this tendency continues to $n_{1/2}$. As will be shown below, the two cancel each other almost completely at $n\approx n_{1/2}= 2 n_0$ for $r\gsim 1$ fm. This reproduces qualitatively -- with (\ref{closure}) --  the dropping $E_{sym}$ up to $n_{1/2}$ of the skyrmion model.

\begin{figure}[ht]
\begin{center}
\includegraphics[width=10.0cm]{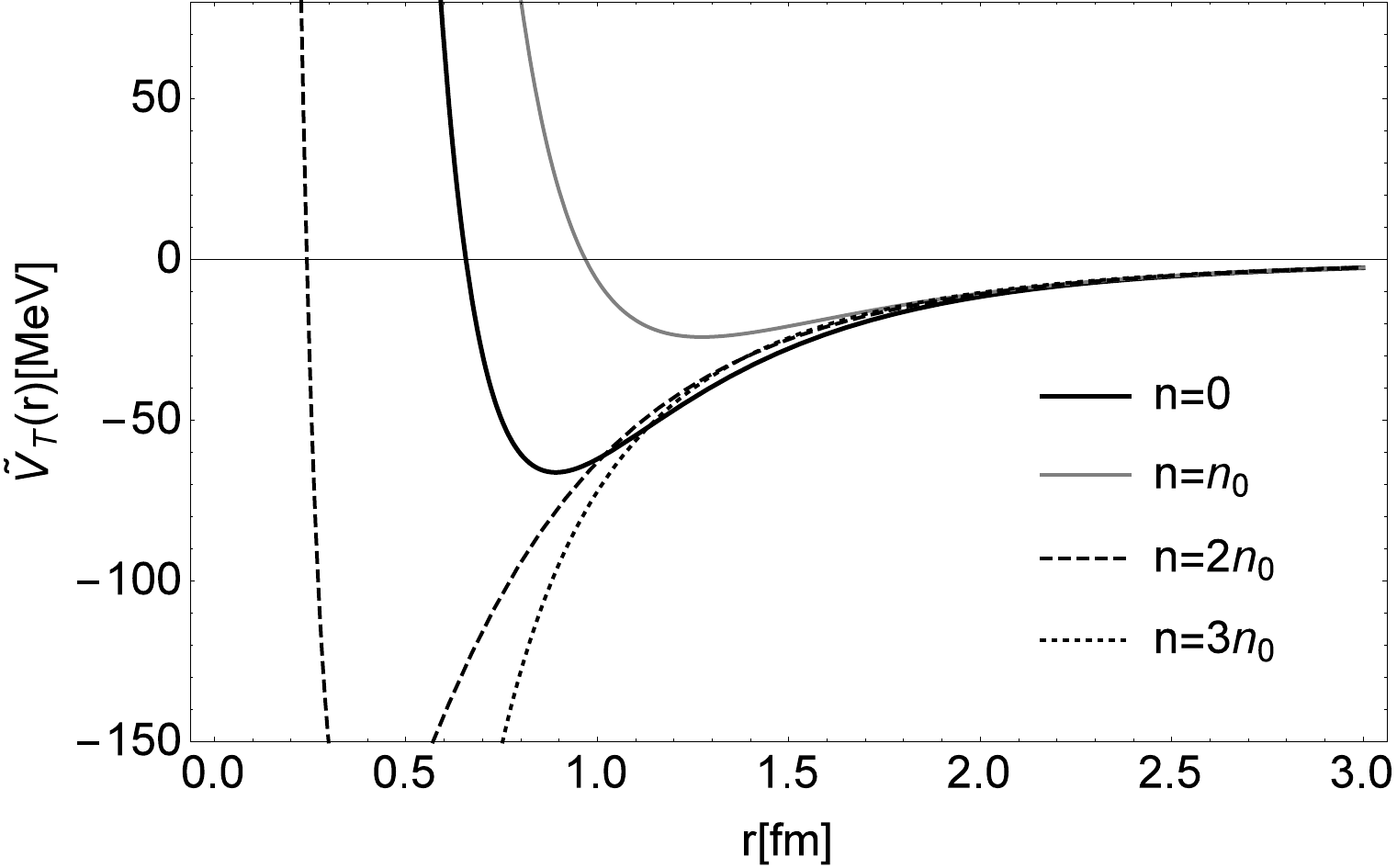}
\caption{$\tilde{V}_T\left(r\right) \equiv V_T\left(r\right) \left( \tau_1\, \tau_2\,S_{12}\right)^{-1}$. For illustration, we take $n_{1/2}=2n_0$.}
%$\Phi_\sigma = \Phi_\rho= 1 - 0.15 \frac{n}{n_0}$ and $\kappa = 1$. }
\label{tensorF}
\end{center}
\end{figure}

Now the only possible way that the cusp structure can arise at $n_{1/2}$ is that the ratio ${\cal R}$ must change drastically. It should drop  faster than the $\rho$ mass does. Here enters the most important ingredient of the theory, that is, the ``vector manifestation (VM)" predicted in hidden local symmetry~\cite{HLS-loops}, that as $\la\bar{q}q\ra\to 0$, the $\rho$ mass goes to the (VM) fixed point,
\be
m_\rho^\ast\propto g^\ast\to 0.\label{rhomass}
\ee
Here $g$ is the hidden gauge coupling constant. With the nucleon mass going to $m_0$, the ratio ${\cal R}$ goes as
\be
{\cal R}\propto (g^\ast/g)^2.\label{RinII}
\ee
With the ratio of the coefficient of the $\rho$ tensor (\ref{RinII}) decreasing faster than the $\rho$ mass,  the $\rho$ tensor gets strongly suppressed for $n\gsim n_{1/2}$. This is shown in Fig.~\ref{tensorF}.
At $n\approx n_{1/2}$ the $\rho$ tensor is more or less completely killed, with the pion tensor taking over, making the net tensor force increase with density. This reproduces the cusp. Thus the implementation of the topology change in the skyrmion crystal model, presumably valid at high density and in the large $N_c$ limit, in $bs$HLS Lagrangian determines the crucial properties of the Lagrangian for $n\gsim n_{1/2}$.

\begin{figure}[ht]
\begin{center}
\includegraphics[width=9cm]{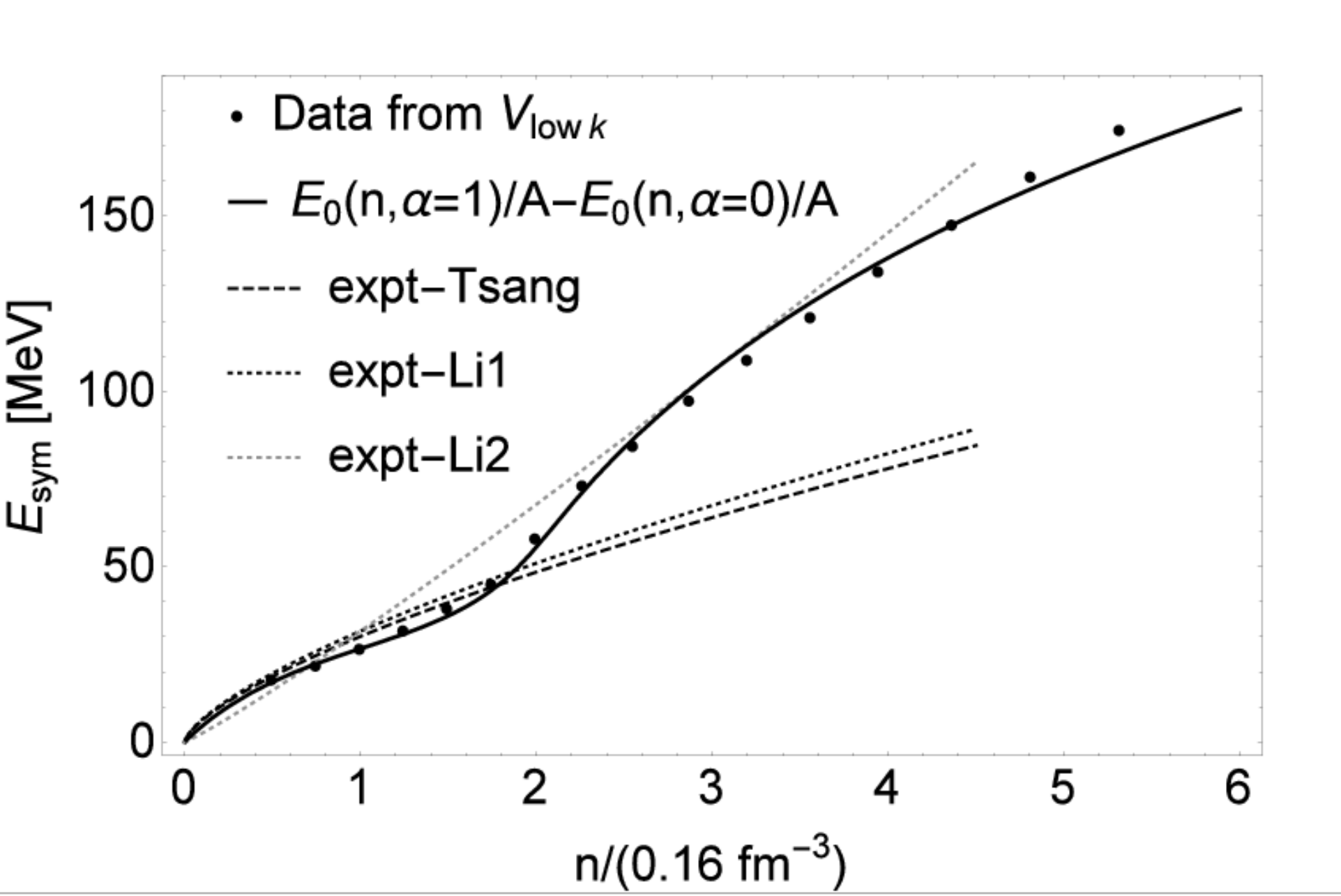}
\caption{The symmetry energy $E_{sym}$.  The constraints presently available  coming from experiments (heavy ions) are indicated. The predicted values are consistent with the data.
 }\label{Esym}
 \end{center}
\end{figure}

What is done up to here is essentially doing a mean-field treatment of the EFT. In the $V_{lowk}$RG framework, the second decimation taking into higher correlations has to be implemented. As shown in \cite{PKLMR}, a full RG analysis in the given framework ``smooths" the cusp structure, giving the changeover from a soft to hard symmetry energy as shown in Fig.~\ref{Esym}.

\begin{figure}[ht]
\begin{center}
\includegraphics[width=10.0cm]{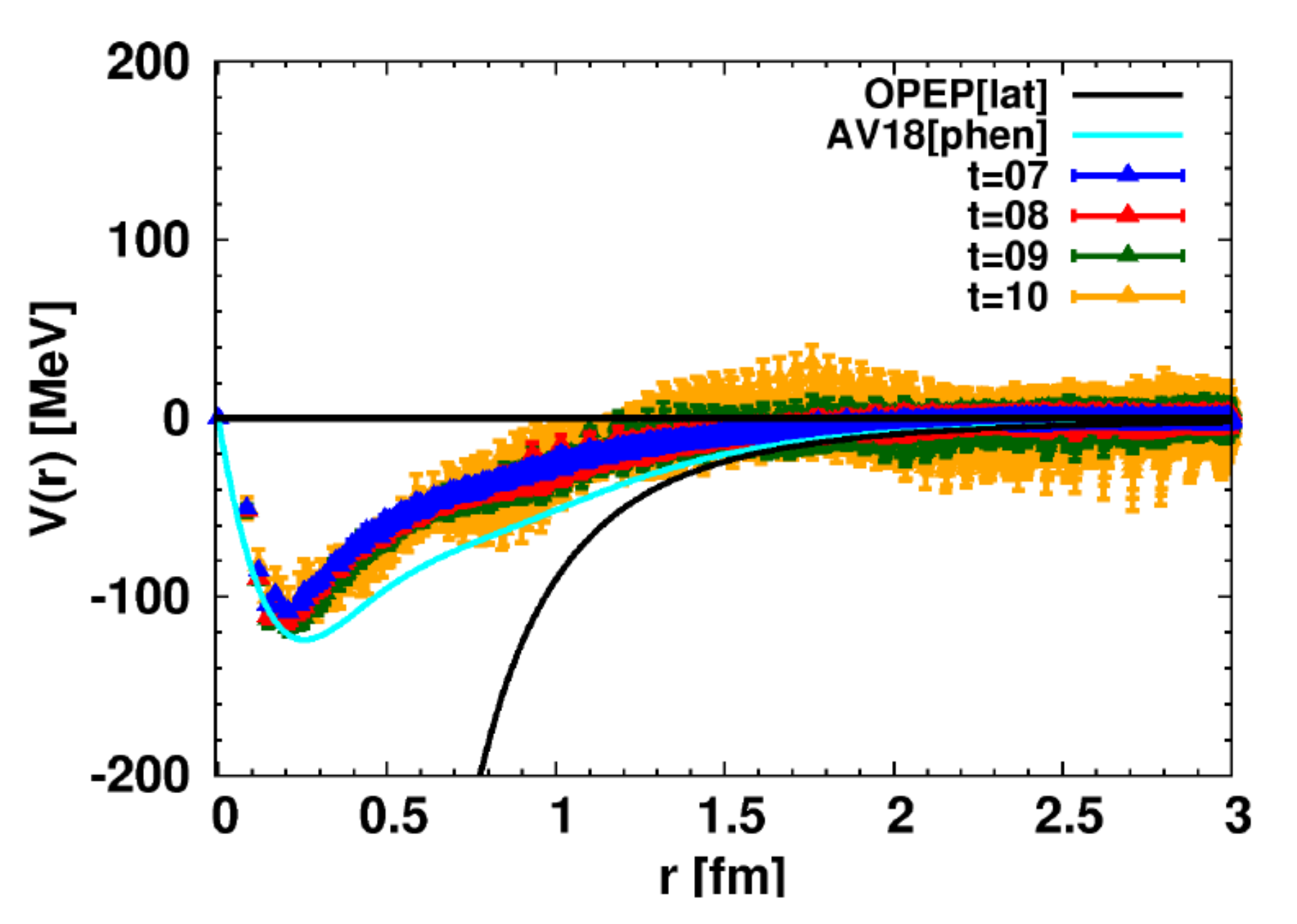}
\caption{Tensor force obtained in lattice QCD~\cite{Doi}. The solid line corresponds to one-pion exchange.  The phenomenological potential (AV18) and the $\rho$/$\pi$-exchange potential are similar.}
\label{lattice-tensor}
\end{center}
\end{figure}

Let me elaborate here on the structure of the net tensor force, which is the key ingredient of  HLS. That the effective tensor force in nuclear interactions is given by the exchanges of $\pi$ and $\rho$ is characteristic of the $bs$HLS formulation. That the nuclear tensor force has a component that is pionic and some other which is not pionic is actually seen in lattice QCD calculations as shown in Fig.~\ref{lattice-tensor}.  The short-range components in the lattice measurement may require contributions from the infinite tower of $\rho$ mesons as in holographic QCD, but it is clear that the lowest-lying $\rho$ should play a dominant role in counter-balancing the pionic tensor in nuclear matter. (This specially important role of the lowest $\rho$ in the nucleon structure has been beautifully confirmed in the holographic dual QCD model anchored on gravity-gauge duality in string theory~\cite{multifacet}.)
It is the delicate interplay between the two tensor forces with the important role of IDD that makes it different from the S$\chi$EFT approach to the problem. In the latter, what corresponds to the $\rho$ tensor would be simulated at higher orders in pionic interaction in chiral perturbation series.  In principle there should be no difficulty in accounting for the tensor force structure in nuclei with pions-only chiral perturbation scheme. This point is discussed in \cite{holtetal}. It is in going to density $n\gsim 2n_0$ that the present approach has an advantage over the S$\chi$EFT.  The $bs$HLS offers a lot simpler accounting of the transition region from  hadronic structure to non-ordinary, be that in terms of topology or hadron-quark continuity.\footnote{I should mention here a difference in opinion on the role of the $\rho$ meson in nuclear effective field theory between Gerry Brown and Steven Weinberg. Gerry insisted on having the $\rho$ degree of freedom explicitly present in nuclear physics, whereas Weinberg did not see the necessity of introducing other degrees of freedom than pions. Having worked for a long time with Gerry, I of course prefer Gerry's point of view. This is the line adopted in this note. There are still lots of mysterious things about HLS that are not understood, such as for instance, the exactness of the KSRF relation and other low-energy theorems and the validity of the VM fixed point and the possibility of  Weinberg's ```mended symmetries" realized with local gauge fields etc. As they stand, Weinberg's view with pions only, I think, seems more ``rigorous" with respect to the FT.}
\subsection{Massive compact stars}
The effective field theory $bs$HLS that incorporates both hidden scale invariance and hidden local symmetry as formulated in terms of $V_{lowk}$RG {\it with only one parameter} describing the scaling $\Phi=f_\chi^\ast/f_\chi\approx f_\pi^\ast/f_\pi$ gives equally well as sophisticated energy-functional  models with $\sim 10$ or more arbitrary parameters do all the properties of normal nuclear matter, such as binding energy, equilibrium density, compression modulus etc. This is not surprising because it is essentially Walecka relativistic mean-field model with multi-dimension  meson fields incorporated, the difference being that here the IDDs encode the multi-dimension fields in mean field. That the mean-field treatment of $bs$HLS-type Lagrangian leads to a Landau Fermi-liquid fixed point theory  and hence is equivalent to Waleck's nonlinear mean-field model was explicitly worked out in Chaejun Song's thesis work~\cite{song}. It was first shown there that having the IDDs in the Lagrangian, if done correctly, does not violate thermodynamic consistency as had been thought incorrectly by workers who were putting density dependence arbitrarily in relativistic mean field theory.

What is not at all trivial is that one {\it also} gets straightforwardly the maximum neutron mass $\sim 2M_\odot$ and radius $R\approx 12$ km, all consistent with the observation.
\begin{figure}[ht]
\begin{center}
\includegraphics[height=6.5cm]{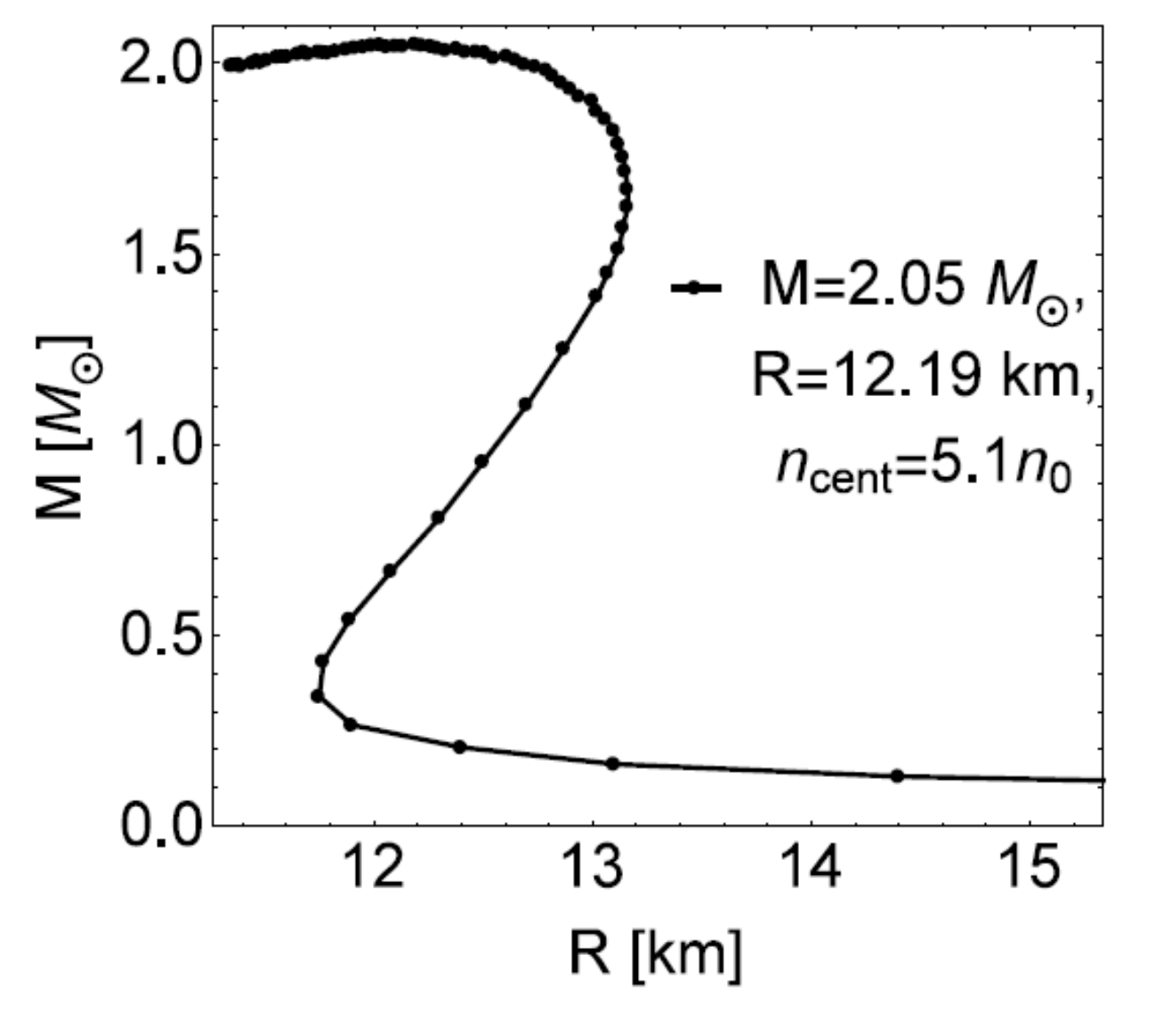}%\includegraphics[height=6cm]{nc.pdf}
\caption{ The mass vs. radius of the neutron star in beta equilibrium.
 }\label{mass-radius}
 \end{center}
\end{figure}
The result of the analysis made in \cite{PKLMR} is depicted in Fig.~\ref{mass-radius}. The large mass results due to the stiffening of the symmetry energy -- and consequently the EoS -- at $n\sim 2n_0$. Here it is the consequence of the topology change, whereas in the description that invokes quark degrees of freedom it is the  hadron-quark crossover at a density $(2-3)n_0$ with strong quark coupling that gives the requisite repulsion~\cite{baym-kojo,fukushima}. This could perhaps be phrased as hadron-quark duality or Cheshire Cat phenomenon.

With large number of parameters, many conventional  nuclear models that may appear different can fit nuclear matter properties and also support massive neutron stars. The problem here is that it is not clear what one is learning. There is little predictiveness in these calculations since the plethora of parameters can fit practically anything. Future experiments, including gravity waves from merging neutron star, may eventually weed out the variety of models and lead to a clearer understanding of the dense matter.

As for the model I am discussing, I must admit there is at present nothing definite to say either that what it describes is closer to nature than any other models as far as the postdictions go. There are however a few {\it predictions with no additional parameters} that are definitely different from other models available in the literature. What's striking, if correct, is that they can bear on the origin of the proton mass, emergent  symmetries in dense matter and their interplay at high density. One striking case is the massive stars' sound velocity $v_s$. The prediction in this theory is that the sound velocity converges to the ``conformal velocity"
\be
v_s^{\rm conformal}=\sqrt{1/3} c.
\ee

To explain what's happening, I need to go a little bit into the issue of scale symmetry and dilaton in QCD. In full generality the matter is quite complex, so I won't go into details. It requires setting up hidden local symmetric Lagrangian including baryons to which the dilaton field is coupled with matter fields as mentioned above with the conformal compensator. There is an unresolved problem of whether one can actually talk about the scalar as a dilaton because the dilaton should be a Nambu-Goldstone boson of spontaneously broken scale symmetry. However it turns out that one cannot talk about spontaneous breaking of scale symmetry without talking about explicit symmetry breaking, which in QCD is due to the trace anomaly. It is not known whether one can therefore have a dilaton treated as a NG boson in QCD with three flavors. There is no lattice calculation that shows that an infrared fixed point exists in QCD for $N_f\leq 3$ relevant for nuclear physics. There have been several attractive schemes proposed~\cite{CT,GS} that allow to set up a scale-chiral perturbative expansion that combines scale symmetry and chiral symmetry with the departure from the presumed IR fixed point as perturbation in scale symmetry breaking on the same order as the quark mass that breaks chiral symmetry.  The problem is that the uncertainty mentioned above on the property of scale symmetry in hadronic physics adds uncertainty in applying the FT to this scale-invariant HLS Lagrangian that we are dealing with, so this represents  an important hole in the reasoning. The procedure adopted -- and its caveats -- in what follows  is explained in \cite{LPR,PKLMR,LMR}.

The $bs$HLS Lagrangian constructed, the details of which are found in \cite{LMR}, is quite involved but it simplifies greatly if one treats the dilaton-matter coupling in the limit that the explicit symmetry breaking is ignored and the explicit symmetry breaking is put entirely in the dilaton potential.  Let me call this ``scale limit." This is somewhat like the chiral limit in which the pseudoscalar NG coupling to matter fields is chiral invariant with the symmetry breaking put  entirely in the NG boson mass term. There is a basic difference between the two limits, however, which makes a caveat to the application of the FT to the dense system we are interested in. The difference is that one can talk about spontaneous breaking of chiral symmetry in the chiral limit, whereas scale symmetry cannot spontaneously break without explicit breaking. Nonetheless I think that the ``scale-limit" treatment that I describe here makes sense. Fortunately what was discussed above and what comes next do not crucially depend on the explicit form of the symmetry breaking potential.

To be as explicit as feasible,  let me write down the relevant part of the Lagrangian from \cite{PKLMR}:
\be
{\cal L}={\cal L}_{\rm inv} +{\cal L}_{\rm SB}  \label{TotalLag}
\ee
where
\begin{eqnarray}
{\cal L}_{\rm inv} &=& {\cal L}_N + {\cal L}_M \, \label{tLag}\\
{\cal L}_N &=&  \bar{N}\left(iD\hspace{-.6em}/
+\frac{{g}_A}{2}\gamma^\mu
\gamma_5 \hat{\alpha}_{\perp\mu} +\frac{{g}_V}{2}\gamma^\mu
 \hat{\alpha}_{\parallel\mu}\right)  N -\zeta m_N\bar{N} N
\label{NucleonLag}    \\
{\cal L}_M &=& \zeta^2 \Big({f_\pi}^2 {\rm
Tr}[\hat{\alpha}_{\perp\mu}\hat{\alpha}_{\perp}^{\mu}] + a {f_\pi}^2{\rm
Tr}[\hat{\alpha}_{\parallel\mu}\hat{\alpha}_{\parallel}^{\mu}]\Big) -
\frac{1}{2g^2}{\rm Tr}[V_{\mu\nu}V^{\mu\nu}] %\nonumber\\
%& &{} + \frac{1}{4} f_\chi^2{\rm Tr}[\hat{\chi} + \hat{\chi}^\dag],
\end{eqnarray}
where the conformal compensator is written with $\zeta=\chi/f_\chi$.
${\cal L}_{\rm SB}$ contains the dilalton potential $V(\chi)$  that breaks scale symmetry, both explicitly and spontaneously, and the chiral symmetry breaking mass term.  For the purpose of what comes below,  we don't need the specific form of the dilaton potential $V(\chi)$. The Maurer-Cartan 1-forms $\hat{\alpha}_\mu$ are made up of covariant derivatives, all defined in \cite{PKLMR}. In \cite{LMR}, the potential is written down explicitly for small scale symmetry breaking.  For ease of notation, the Lagrangian is written in $U(2)$ symmetric form for the vector mesons.

In the $V_{lowk}$RG approach, doing mean-field treatment of the Lagrangian (\ref{TotalLag}) corresponds to doing the first decimation in RG flow. It is a bit involved but the result is rather simple. Since it contains the key points, let me go into some details of the argument involved.

The thermodynamic potential $\Omega = E -TS - \mu N$ at zero temperature  taken in mean-field with (\ref{TotalLag}) for symmetric nuclear matter\footnote{For ease of writing in this part, the $\ast$ is ignored on the in-medium quantities -- density-dependent due to the IDDs -- apart from the ``effective" nucleon mass that gets a tadpole contribution in addition to the IDDs in medium.  No confusion arises for the result we are getting at. For the symmetric nuclear matter in the mean field, the $\rho$ does not contribute. $U(2)$ symmetry is broken at high density $n\gsim n_{1/2}$, therefore the subscript $\omega$ stands for local hidden $U(1)$ symmetry for $\omega$. Explanations for additional notations, which are not essential for the discussion but figure nontrivially in the hidden local symmetry theory, are in order for $g_{V\omega}$ and $f_{\sigma\omega}$. The former is the loop correction in decimating from the matching scale to the scale where the Lagrangian is defined. It is just the loop renormalization to the hidden gauge coupling to the nucleon. The latter is the decay constant of the would-be scalar NG boson that gets ``eaten" by Higgs mechansim to give the mass to $\omega$. All these details are found in \cite{Interplay}.  }
\begin{eqnarray}
\left. \frac{\Omega(T=0)}{V} \right|_{\omega_0 = \langle \omega_0 \rangle,\, \chi = \langle \chi \rangle}
&=& \frac{1}{4\pi^2} \left[ 2E_F^3 k_F - m_N^{\ast\,2}E_F k_F - m_N^{\ast\,4} \ln\left( \frac{E_F + k_F}{m_N^\ast}\right) \right] +  V(\langle \chi \rangle) \nonumber\\
&& + \left[ g_\omega \left( g_{V\omega} -1 \right) \langle \omega_0 \rangle -\mu \right] \frac{2}{3\pi^2} k_F^3 -\frac{1}{2} f_{\sigma \omega}^2 g_\omega^2 \frac{\langle \chi \rangle^2}{f_\sigma^2} \langle \omega_0 \rangle^2\label{Omega}
\end{eqnarray} where $\langle \omega_0 \rangle$ and $\langle \chi \rangle$ are the ``vacuum" (medium) expectation value (VeV) of $\omega_{0}$ and $\chi$, $m_N^\ast = \frac{\la\chi\ra}{f_\sigma} m_N $ and $E_F = \sqrt{k_F^2 + m_N^{\ast\,2}}$. The nucleon number density is
\begin{equation}
n \equiv N/V =  -\frac{\partial (\Omega/V)}{\partial \mu} = \frac{2}{3\pi^2} k_F^3
\end{equation}
where $k_F$ is the Fermi momentum and the chemical potential $\mu$ given by the condition  $\frac{\partial (\Omega/V)}{\partial n} = 0$
is
\begin{equation}
\mu = E_F + g_\omega \left( g_{V\omega} -1 \right) \langle \omega_0 \rangle\,.
\end{equation}
The energy density $\epsilon$ and the pressure $P$ at $T=0$ are given by
\begin{eqnarray}
\epsilon
&=& \frac{1}{4\pi^2} \left[ 2E_F^3 k_F - m_N^{\ast\,2}E_F k_F - m_N^{\ast\,4} \ln\left( \frac{E_F + k_F}{m_N^\ast}\right) \right]  \nonumber\\
&& + g_\omega \left( g_{V\omega} -1 \right) \langle \omega_0 \rangle n -\frac{1}{2} f_{\sigma \omega}^2 g_\omega^2 \frac{\langle \chi \rangle^2}{f_\sigma^2} \langle \omega_0 \rangle^2 +  V(\langle \chi \rangle) \label{eden}
\end{eqnarray}
and
\begin{eqnarray}
P &=& -\left. \frac{\Omega}{V} \right|_{\omega_0 = \langle \omega_0 \rangle,\, \chi = \langle \chi \rangle} \\
&=& \frac{1}{4\pi^2} \left[ \frac{2}{3}E_F k_F^3 - m_N^{\ast\,2}E_F k_F + m_N^{\ast\,4} \ln\left( \frac{E_F + k_F}{m_N^\ast}\right) \right] \nonumber\\
&& + \frac{1}{2} f_{\sigma \omega}^2 g_\omega^2 \frac{\langle \chi \rangle^2}{f_\sigma^2} \langle \omega_0 \rangle^2 -  V(\langle \chi \rangle)\,.\label{pre}
\end{eqnarray}
From the stationarity  conditions for the gap equations for $\chi$ and $\omega$
\begin{eqnarray}
\left. \frac{\partial\, \Omega }{\partial \chi} \right|_{\omega_0 = \langle \omega_0 \rangle,\, \chi = \langle \chi \rangle} = 0\,, \quad \left. \frac{\partial \,\Omega }{\partial \omega_0} \right|_{\omega_0 = \langle \omega_0 \rangle,\, \chi = \langle \chi \rangle} = 0\,
\end{eqnarray}
we have
\begin{eqnarray}
&\frac{m_N^2\langle \chi \rangle}{\pi^2 f_\sigma^2} \left[ k_F E_F - m_N^{\ast\,2} \ln \left( \frac{k_F + E_F}{m_N^\ast} \right) \right] -\frac{f_{\sigma\omega}^2}{f_\sigma^2} g_\omega^2 \langle \omega_0\rangle^2 \langle \chi \rangle + \left. \frac{\partial\, V(\chi)}{\partial \chi} \right|_{\chi = \langle \chi \rangle} =0 \,, \label{gap1}\\
&g_\omega \left(g_{V\omega}-1 \right)n -f_{\sigma \omega}^2 g_\omega^2 \frac{\langle \chi \rangle^2}{f_\sigma^2} \langle \omega_0 \rangle = 0\,. \label{gap2}
\end{eqnarray}
One obtains from (\ref{gap1}) and (\ref{gap2}) the VeV of the trace of energy-momentum tensor $\theta_\mu^\mu$, which in the chiral limit reads
\begin{eqnarray}
\langle \theta^\mu_\mu \rangle
&=& \langle \theta^{00} \rangle - \sum_i \langle \theta^{ii}\ra = \epsilon - 3 P\nonumber\\
& =& 4V(\langle \chi \rangle) - \langle \chi \rangle \left. \frac{\partial V( \chi)}{\partial \chi} \right|_{\chi = \langle \chi \rangle}. \label{TEMT1}
\ee
This is just what one gets by taking in the chiral limit the mean-field value of the TEMT $\theta_\mu^\mu$,
\begin{equation}
\theta_\mu^\mu= 4V(\chi) - \chi \frac{\partial V(\chi)}{\partial \chi} . \label{TEMT0}
\end{equation}
Thus in the mean field of $bs$HLS, the trace of energy momentum tensor (TEMT) is given solely by the dilaton condensate.
What this shows is that the Fermi surface does not spoil scale symmetry.

%Furthermore it has been verified that strongly-correlated hadronic interactions that can come in in the second decimation do not modify the dilaton potential.
%Thus if the dilaton condensate is a constant of, or depends weakly on,  density,  then $\frac{\del}{\del n} \TEMT$ will be zero or approximately zero.  This feature will account for the emergent scale symmetry in compact-star matter.
Now a most striking observation in the mean-field treatment is that the nucleon mass scales in density as
\be
m_N^\ast/m_N\approx \la\chi\ra^\ast/\la\chi\ra
\ee
and that for a specific scaling of the renormalization constant $(g_{V\omega}-1)$ { the dilaton condensate is independent of density}. This is shown in Fig.~\ref{mass-coupling} taken from \cite{Interplay}.
\begin{figure}[h]
\begin{center}
\includegraphics[width=10cm]{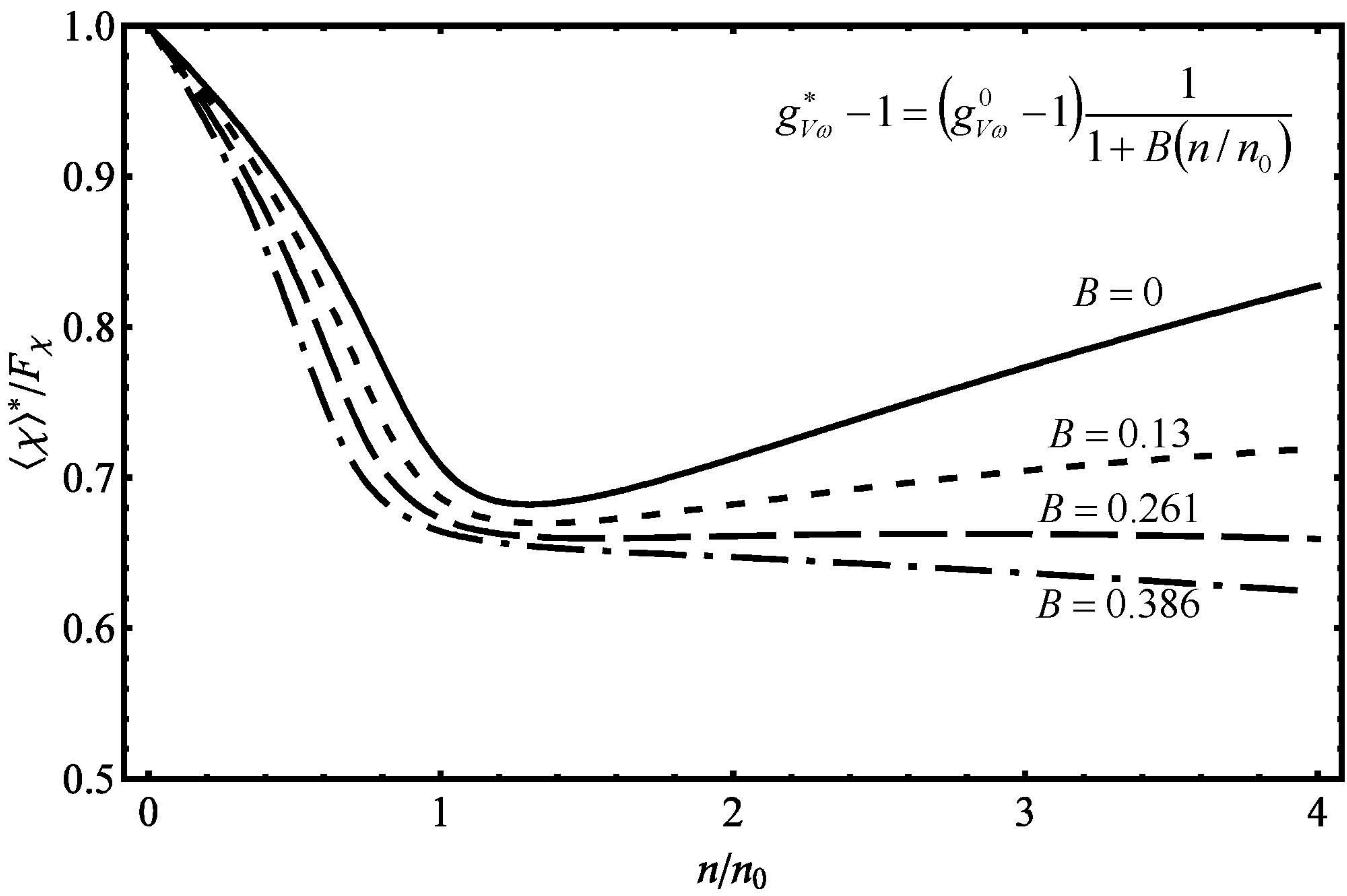}
\caption{The ratio $m_N^*/m_N\approx \la\chi\ra^*/\la\chi\ra_0$  as a function of
density for varying density dependence of $g_{V\omega}^\ast$.
What is notable is that the nucleon mass stops dropping at a density slightly
above nuclear matter density $n_0$ and stays more or less constant above that
density.
}
\label{mass-coupling}
\end{center}
\end{figure}
{\it What this means is that at some higher density,  the trace of energy momentum tensor, a non-zero value, is independent of density.}
This observation agrees with the parity-doubling scenario for the nucleon~\cite{paritydoublet}. This result will be supported in the $V_{lowk}$RG calculation going beyond the mean-field as will be shown below. An important point to make at this point is that {\it parity doubling, absent in low-density regime, emerges in this theory at $n_{1/2}$ with the topology change.} This is related also to the emergence of scale symmetry in nuclear medium in the guise of ``dilaton-limit fixed point (DLFP)"~\cite{beane,SLPR,PKLMR}. In the parity-doublet model, the chiral-invariant mass $m_0$ is put in by hand. Here a mass $m_0^\prime \propto \la\chi\ra$ emerges from strong correlations in hadronic interactions.

The mean-field treatment is a single-decimation approach to Landau Fermi-liquid fixed point theory. It is the limit $N\to \infty$ where $N=k_F/(\Lambda-k_F)$ where $\Lambda$ is the cut-off scale for the decimation. The $V_{lowk}$RG does better in the sense that it takes into account certain $1/N$ corrections via ring diagrams~\cite{ringdiagram,siu09}.

Unlike the mean-field calculation, it is not feasible to do analytic calculation in $V_{lowk}$RG. It is purely numerical. There is one intriguing complication in doing the numerical work that is absent in the mean-field calculation. In the mean-field calculation of the TEMT, all matter field contributions drop out, that is,  they get cancelled away exactly, leaving terms involving only the diilaton field. The IDDs do not figure in the TEMT. A similar cancelation takes place in $V_{lowk}$RG, however,  for a certain condition on the vector manifestation with the $\rho$ meson as described below.

The result of the calculation of the TEMT in $V_{lowk}$RG with $bs$HLS Lagrangian carrying IDDs~\cite{PKLMR} as described above is given in Fig.~\ref{TEMT}.
\begin{figure}[h]
\begin{center}
\includegraphics[width=10cm]{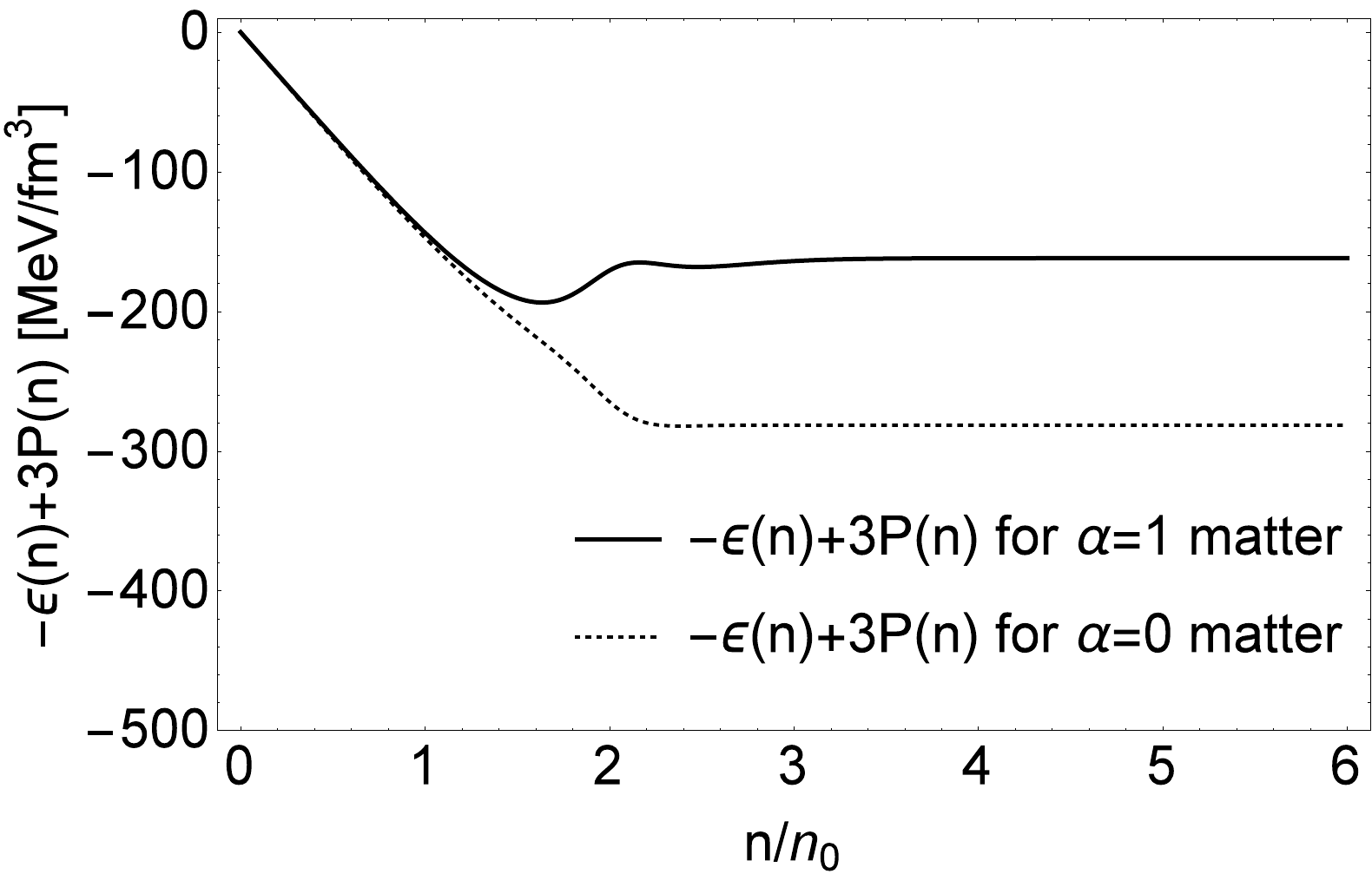}
\caption{ $\la\theta_\mu^\mu\ra$  for both nuclear and pure neutron matter.
 }\label{TEMT}
 \end{center}
\end{figure}
What comes out in this high-order calculation turns out to be remarkably close to the mean-field  result. This indicates that the cancellation that takes place in the mean field persists at higher orders. This is consistent with the structure of the Lagrangian that has the explicit scale symmetry breaking lodged entirely in the dilaton potential. Neither the Fermi sea nor high-order nuclear correlations disturb it. Also the strength of the explicit symmetry breaking does not influence the feature that the TEMT is independent of density. This is important for the sound velocity.
\subsection{Sound velocity $v_s=\frac{1}{\sqrt{3}}c$}
The density independence of the TEMT for $n\gsim 2n_0$ predicts a precocious onset of conformal sound velocity in massive compact stars~\cite{PR-sound,PKLMR}.
\begin{figure}[h]
\begin{center}
\includegraphics[width=10cm]{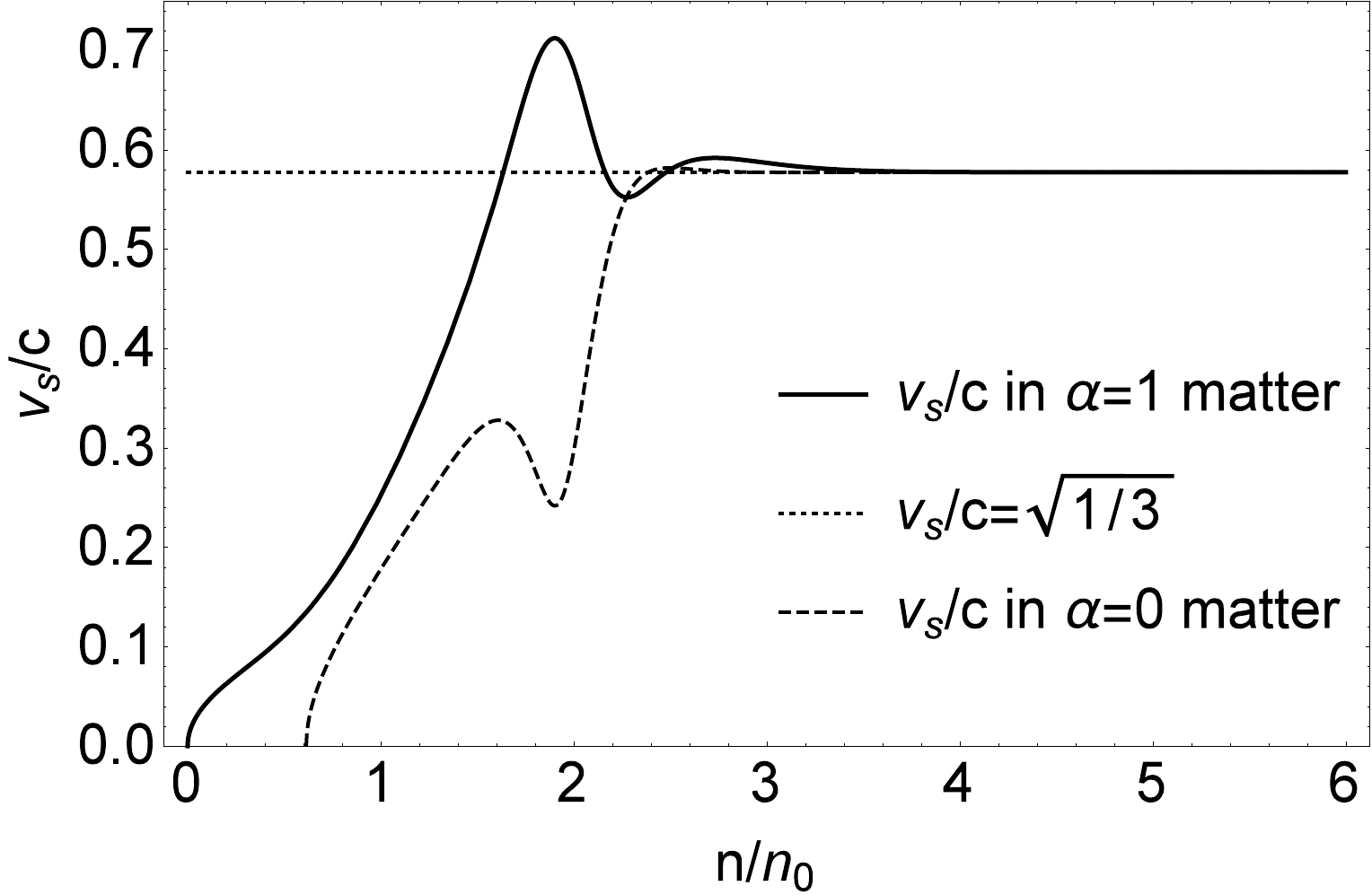}
\caption{Sound velocity for both nuclear and pure neutron matter.
 }\label{SV}
 \end{center}
\end{figure}
To see this, look at the relation between the trace of energy-momentum tensor and the sound velocity, which is given by
\be
\frac{\partial}{\partial n} \la\theta_\mu^\mu\ra=\frac{\partial \epsilon (n)}{\partial n} \Big(1-3\frac{v_s^2}{c^2}\Big)\label{derivTEMT}
\ee
with $v_s^2/c^2=\frac{\partial P(n)}{\partial n}/\frac{\partial \epsilon(n)}{\partial n}$. Since  we have, from Fig.~\ref{TEMT}, the TEMT  independent of density, the left-hand side of (\ref{derivTEMT}) is  zero.  If we assume that there is no extremum in the energy density for density $n > n_{1/2}$ -- which we believe is justified\footnote{So far there is no indication for abnormal states of the type called ``Lee-Wick state"  proposed  by Lee and Wick~\cite{lee-wick}.},  then $\frac{\partial \epsilon (n)}{\partial n} \neq 0$. It then follows that
\be
v_s^2/c^2\approx 1/3.
\ee
One can see in Fig.~\ref{SV} that this indeed is verified in the calculation.

It is interesting to see how the conformal velocity $(1/\sqrt{3})c$ \index{conformal velocity} is approached. It starts with $v_s^2/c^2 < 1/3$ at low density $n <2n_0$, goes up to $v_s^2/c^2 > 1/3$ at $n\sim 2n_0$, drops below $1/3$ and then climbs to, and asymptotes at, 1/3 at higher density $n\gsim 3n_0$.  This feature closely resembles the scenario arrived at by Bedaque and Steiner~\cite{bedaque} in the study of the sound velocity based on their analysis of neutron stars with mass around two solar mass with various phenomenological equations of state. In our theory, there is a rather abrupt changeover of the $bs$HLS Lagrangian parameters due to a topology change (i.e.,  the cusp in the skyrmion crystal description), so one might imagine that such a behavior could be an artifact of the sharp transition. One of the characteristic features arising from the topology change, as stressed,  is the stiffening of the symmetry energy at higher density $n > 2n_0$.  It is responsible for the relatively high  proton fraction of nuclear matter in beta equilibrium. It might render the direct URCA process to set in precociously and trigger too rapid a star cooling which might be at odds with observation. If it turned out to be serious,  then that would indicate within our formalism that we need to improve on how the vector manifestation property of the $\rho$ meson sets in at high density.
%This matter may be related to what's mentioned in footnote 10 regarding the effective IDD of the gauge coupling constant.
On the other hand, the fact that the dense baryonic matter in our description is in the precursor state to an emerging scale invariance and hence manifesting a conformal-type sound velocity as we are proposing is highly suggestive of the intricate mechanism of the Bedaque-Steiner scenario. What is surprising in our description is, however,  that what attributes to ``conformality" with  $v_s^2/c^2=1/3$ sets in so precociously in density and for $\theta_\mu^\mu\neq 0$. We find it appealing to identify this phenomenon as a precursor to an emergent conformality. In \cite{bedaque}, in contrast,  the matter with $v_s^2/c^2 > 1/3$ should prevail up to density $n \sim 5n_0$, the maximum central density of $\sim 2$-solar mass objects because of the strong hadronic interactions intervening in the phenomenological models they relied on. As noted before, all phenomenological models fit accurately to the properties of nuclear matter (at $n=n_0$) and consistent with the symmetry ``constraints" have $v_s^2/c^2 > 1/2$~\cite{haensel}. It is of course expected that the conformal velocity will  appear at very super-high density.
\section{Discussions}
While the first part of this article dealt with what I would consider as a rather convincing ``proof" of the FT -- and its corollary,  the second part went into an uncharted domain with certain unverifiable guess works. However some of the results are quite reasonable and at the same time intriguing. The intriguing part has to do with the sound velocity with the compact stars. In the literature there is practically no discussion of what a particular sound velocity implies other than the constraint due to causality, which forbids the sound velocity from exceeding the velocity of light. The $V_{lowk}$RG calculation that correctly describes normal nuclear systems, both finite and infinite, and also accounts well for the observed massive compact stars predicts the emergence at high density of various symmetries, party doubling, scale symmetry, hidden local symmetry etc. invisible in  the matter-free vacuum. At present those features have not been confirmed or invalidated by Nature.  This theory predicts that the sound velocity of  massive stars approaches the conformal velocity $(1/\sqrt{3})c$ at a density $\sim 3 n_0$ and stays the same up to the maximum density supported by the star $\sim (5-6) n_0$. The tendency for the sound velocity to shoot up with stiff EoS seen in conventional nuclear models is absent in this description even though the symmetry energy is stiffened due to the topology change and accommodates the massive stars.

What is interesting about this prediction is that the VeV of the trace of energy-momentum tensor $\la\theta_\mu^\mu\ra$ is independent of density for $n\gsim n_{1/2}\sim 2n_0$. This means that in the RG flow, both the IDDs inherited from QCD and the induced density dependence due to  nuclear correlations must be canceling exactly in the TEMT. The cancelation takes place both in the single-decimation calculation and  in the double decimation calculation. This means that the cancelation takes place density-independently.

Another puzzling observation is that in the $V_{lowk}$RG, there is an intricate interplay of the $\omega$-nucleon coupling and the vector manifestation of the $\rho$ meson. The former is seen at the mean-field level as in Fig.~\ref{mass-coupling} and the latter requires a large density at which the VM manifestation sets in.  In \cite{PKLR}, the VM took  place at $n\approx 6.6n_0$ and  the sound velocity did not converge to the conformal value, but increased monotonically beyond $(1/\sqrt{3})c$.  This  means that the density dependence coming from the IDDs and the correlation-induced effect remain uncanceled.   In \cite{PKLMR}, the VM fixed point is required to be at $\sim 25n_0$, which then makes the sound velocity converge, way before the VM fixed point,  to the conformal value. Other than the VM fixed point, the two calculations are qualitatively  -- and semi-quantitatively -- the same. This shows that the VM fixed point density plays a crucial role for the sound velocity.

In discussing the Hoyle state,  I alluded at Cheshire Cat  for a possible link between the description of an {\it ab initio} approach with S$\chi$EFT potential and a skyrmion description where topology and strong nuclear correlations could be trading in.  There also seems to insinuate a similar duality between the description of massive compact stars based on the  skyrmion-half-skyrmion topology change described in this note  and the one anchored on the hadron-quark (e.g., quarkyonic) crossover discussed in \cite{fukushima,baym-kojo}, both taking place at density $\sim 2 n_0$. It is not at all obvious how to formulate the suggestive connection as in the case treated in terms of the chiral bag and skyrmions for, e.g., the flavor singlet axial charge worked out by Hee-Jung Lee and collaborators~\cite{hjlee-fsac}, but it would be worth thinking about it.

Finally I must mention a crucial open problem that needs to be resolved to confirm or invalidate the scenario developed for compact-star matter.   The basic assumption made throughout is that the explicit scale symmetry breaking can be ignored in the matter sector, with the symmetry breaking effect lodged entirely in  the dilaton potential.  As shown in \cite{LMR}, this is justified in the scale-chiral expansion if the anomalous dimension of $G^2_{\mu\nu}$ -- denoted $\beta^\prime$ -- where $G_{\mu\nu}$ is the gluon stress-energy tensor satisfies $|\beta^\prime|\ll 1$.  At present this quantity is totally unknown for QCD with $N_f\sim 3$. It could be $\sim 3$ or bigger~\cite{shrock}. Intriguingly there is an indication that a skyrmion description of dense matter, realistically formulated, could provide an answer~\cite{hWZ}.

\subsection*{Acknowledgments}
I am deeply grateful for many years' helpful and enlightening discussions with Gerry Brown, Jean Delorme, Bengt Friman, Masa Harada, Kuniharu Kobodera, Tom Kuo, Hyun Kyu Lee, Yong-Liang Ma, Dong-Pil Min, Byung-Yoon Park,  Tae-Sun Park, Won-Gi Paeng, Vicente Vento,  Denys Wilkinson and Ismael Zahed. I would like to thank also Rod Crewther, Lewis Tunstall and Koichi Yamawaki for informing me of their views on scale symmetry and the IR fixed point in QCD.

\end{document}